\documentclass[aps,preprint]{revtex4}%
\usepackage{amsfonts}
\usepackage{amsmath}
\usepackage{amssymb}
\usepackage[colorlinks,linkcolor=blue,citecolor=blue,urlcolor=blue]{hyperref}
\usepackage{subfigure}
\usepackage{graphicx}
\usepackage{cleveref}
\usepackage{listings}
\usepackage{xcolor}
\usepackage{array}
\usepackage{url}%
\begin{document}
\preprint{CTP-SCU/2020037}
\title{Scalarized Einstein-Born-Infeld-scalar Black Holes }
\author{Peng Wang}
\email{pengw@scu.edu.cn}
\author{Houwen Wu}
\email{iverwu@scu.edu.cn}
\author{Haitang Yang}
\email{hyanga@scu.edu.cn}
\affiliation{Center for Theoretical Physics, College of Physics, Sichuan University,
Chengdu, 610064, China}

\begin{abstract}
The phenomenon of spontaneous scalarization of Reissner-Nordstr\"{o}m (RN)
black holes has recently been found in\ an Einstein-Maxwell-scalar (EMS) model
due to a non-minimal coupling between the scalar and Maxwell fields.
Non-linear electrodynamics, e.g., Born-Infeld (BI) electrodynamics,
generalizes Maxwell's theory in the strong field regime. Non-minimally
coupling the BI field to the scalar field, we study spontaneous scalarization
of an Einstein-Born-Infeld-scalar (EBIS) model in this paper. It shows that
there are two types of scalarized black hole solutions, i.e., scalarized
RN-like and Schwarzschild-like solutions. Although the behavior of scalarized
RN-like solutions in the EBIS model is quite similar to that of scalarize
solutions in the EMS model, we find that there exist significant differences
between scalarized Schwarzschild-like solutions in the EBIS model and
scalarized solutions in the EMS model. In particular, the domain of existence
of scalarized Schwarzschild-like solutions possesses a certain region, which
is composed of two branches. The branch of larger horizon area is a family of
disconnected scalarized solutions, which do not bifurcate from scalar-free
black holes. However, the branch of smaller horizon area may or may not
bifurcate from scalar-free black holes depending on the parameters.
Additionally, these two branches of scalarized solutions can be both
entropically disfavored over comparable scalar-free black holes in some
parameter region.

\end{abstract}
\maketitle
\tableofcontents

\section{Introduction}

Recent observations of gravitational waves \cite{Abbott:2016blz} and the
black-hole shadow \cite{Akiyama:2019cqa} have provided us with a better
understanding of black hole nature in the strong gravity regime. Specially,
these observations have opened up a new era to test the no-hair theorem
\cite{Israel:1967wq,Carter:1971zc,Ruffini:1971bza,Khodadi:2020jij}, which is
crucial to understand black hole physics. The no-hair theorem states that a
black hole is uniquely characterized by three observable properties, its mass,
angular momentum and electrical charge. Testing the no-hair theorem in
experiments can help us constrain alternative theories of gravity, which often
contain extra scalar fields. On the other hand, black holes can become a new
venue for the detection of the scalar fields themselves
\cite{Brito:2015oca,Chen:2019fsq}. Although the no-hair theorem has been
proven for a number of field-theory models (e.g., the Einstein-Maxwell
theory), various counter-examples were found. The first counter-example is
hairy black hole solutions\ in\ the context of the Einstein-Yang-Mills theory
\cite{Volkov:1989fi,Bizon:1990sr,Greene:1992fw}. Later, black hole solutions
with Skyrme hairs \cite{Luckock:1986tr,Droz:1991cx} and dilaton hairs
\cite{Kanti:1995vq} were obtained. For a review, see \cite{Herdeiro:2015waa}.

To understand the formation of hairy black holes, the phenomenon of
spontaneous scalarization was discussed for neutron stars in scalar-tensor
models \cite{Damour:1993hw}. In this phenomenon, the non-minimal coupling of
the scalar field to the Ricci curvature can lead to a certain parameter
region, where scalar-free and scalarized neutron star solutions coexist, and
the scalarized solution is energetically favoured. Later on, spontaneous
scalarization was generalized to black holes in scalar-tensor models
\cite{Cardoso:2013opa,Cardoso:2013fwa}. Recently, spontaneous scalarization
has been realized in the extended Scalar-Tensor-Gauss-Bonnet (eSTGB) gravity,
in which the scalar field is non-minimally coupled to Gauss-Bonnet curvature
corrections of the gravitational sector
\cite{Doneva:2017bvd,Silva:2017uqg,Antoniou:2017acq,Doneva:2018rou,Cunha:2019dwb,Herdeiro:2020wei}%
. However in the eSTGB models, studying dynamical evolution is challenging due
to non-linear curvature terms in the evolution equations. Subsequently, a
simpler type of model, namely the Einstein-Maxwell-scalar (EMS) models, was
introduced to gain a deeper insight into spontaneous scalarization
\cite{Herdeiro:2018wub}. In the EMS models, spontaneous scalarization can be
triggered by the strong non-minimal coupling of the scalar field to the
electromagnetic field. A massless and non-self-interacting scalar field was
considered in \cite{Herdeiro:2018wub}, where an exponential coupling function
was introduced to ensure a tachyonic instability of Reissner-Nordstr\"{o}m
(RN) black holes. Shortly afterwards, spontaneous scalarization in the EMS
models was further discussed in context of coupling functions beyond the
exponential coupling \cite{Fernandes:2019rez,Blazquez-Salcedo:2020nhs}, dyons
including magnetic charges \cite{Astefanesei:2019pfq}, axionic-type couplings
\cite{Fernandes:2019kmh}, massive and self-interacting scalar fields
\cite{Zou:2019bpt,Fernandes:2020gay}, horizonless reflecting stars
\cite{Peng:2019cmm}, linear stability of scalarized black holes
\cite{Myung:2018vug,Myung:2019oua,Zou:2020zxq}, higher dimensional scenario
\cite{Astefanesei:2020qxk}, quasinormal modes of scalarized black holes
\cite{Myung:2018jvi,Blazquez-Salcedo:2020jee}, two U(1) fields
\cite{Myung:2020dqt}, and quasi-topological electromagnetism
\cite{Myung:2020ctt}. Analytic approximations were also used to study
spontaneous scalarization of the EMS models
\cite{Konoplya:2019goy,Hod:2020ljo,Hod:2020ius}.

Non-linear electrodynamics (NLED) is an effective model incorporating quantum
corrections to Maxwell's electromagnetic theory. In the Einstein-NLED
theories, various NLED charged black holes were derived and discussed in a
number of papers
\cite{Soleng:1995kn,AyonBeato:1998ub,Maeda:2008ha,Guo:2017bru,Mu:2017usw,Wang:2019kxp,Wang:2019jzz,Wang:2018hwg}%
. Among various NLED, there is a famous string-inspired one: Born-Infeld (BI)
electrodynamics, which encodes the low-energy dynamics of D-branes. BI
electrodynamics was first proposed to smooth divergences of the electrostatic
self-energy of point charges by introducing a cutoff on electric fields
\cite{Born:1934gh}. After BI black hole solutions were obtained
\cite{Dey:2004yt,Cai:2004eh}, their properties have been extensively
investigated in the literature
\cite{Fernando:2003tz,Stefanov:2007eq,Banerjee:2010da,Doneva:2010ke,Zou:2013owa,Hendi:2015hoa,Zeng:2016sei,Li:2016nll,Tao:2017fsy,Dehyadegari:2017hvd,Wang:2018xdz,Liang:2019dni,Gan:2019jac}%
.

One of the motivations for this work is to understand the effect of NLED
corrections on the EMS models. Specifically, we numerically obtain and study
scalarized black hole solutions in the Einstein-Born-Infeld-scalar (EBIS)
model with an exponential coupling. The rest of this paper is organized as
follows. Section \ref{Sec:EBISM} presents the basics of the EBIS model and
provides the equations of motion for the solution ansatz of interest. In
section \ref{Sec:BIBH}, we discuss various properties of scalar-free BI black
hole solutions. Section \ref{Sec:SBIBH} contains our main numerical results
for scalarized black hole solutions, which include domains of existence,
thermodynamic preference and effective potentials for radial perturbations. We
summarize our results with a brief discussion in section \ref{Sec:DC}.

\section{EBIS Model}

\label{Sec:EBISM}

We consider a real scalar field $\phi$ minimally coupled to Einstein's gravity
and non-minimally coupled to the BI electromagnetic field $A_{\mu}$, which is
described by the action%
\begin{equation}
S=\int d^{4}x\sqrt{-g}\left[  R-2\partial_{\mu}\phi\partial^{\mu}\phi
+\frac{4f\left(  \phi\right)  }{a}\left(  1-\sqrt{1+aF^{\mu\nu}F_{\mu\nu}%
/2}\right)  \right]  . \label{eq:Action}%
\end{equation}
Here we take $16\pi G=1$ for simplicity, the coupling parameter $a$ is related
to the string tension $\alpha^{\prime}$ as $a=\left(  2\pi\alpha^{\prime
}\right)  ^{2}>0$, $F_{\mu\nu}=\partial_{\mu}A_{\nu}-\partial_{\nu}A_{\mu}$ is
the BI electromagnetic field strength tensor, and $f\left(  \phi\right)  $ is
the coupling function governing the non-minimal coupling of $\phi$ and
$A_{\mu}$. When $a\rightarrow0$, we can recover the EMS model from the action
$\left(  \ref{eq:Action}\right)  $. The equations of motion that follow from
the action $\left(  \ref{eq:Action}\right)  $ are%
\begin{gather}
R_{\mu\nu}-\frac{1}{2}Rg_{\mu\nu}=\frac{\mathcal{T}_{\mu\nu}}{2},\nonumber\\
\partial_{\mu}\left[  \frac{\sqrt{-g}f\left(  \phi\right)  F^{\mu\nu}}%
{\sqrt{1+aF^{\mu\nu}F_{\mu\nu}/2}}\right]  =0,\label{eq:eom}\\
\frac{\partial^{\mu}\left(  \sqrt{-g}\partial_{\mu}\phi\right)  }{\sqrt{-g}%
}=-\dot{f}\left(  \phi\right)  \frac{1-\sqrt{1+aF^{\mu\nu}F_{\mu\nu}/2}}%
{a},\nonumber
\end{gather}
where $\dot{f}\left(  \phi\right)  \equiv df\left(  \phi\right)  /d\phi$, and
the energy-momentum tensor is given by
\begin{equation}
\mathcal{T}_{\mu\nu}=4\left(  \partial_{\mu}\phi\partial_{\nu}\phi
-\frac{g_{\mu\nu}\partial_{\rho}\phi\partial^{\rho}\phi}{2}\right)  +4f\left(
\phi\right)  \left[  \frac{1-\sqrt{1+aF^{\mu\nu}F_{\mu\nu}/2}}{a}g_{\mu\nu
}+\frac{F_{\mu\rho}F_{\nu}^{\text{ }\rho}}{\sqrt{1+aF^{\mu\nu}F_{\mu\nu}/2}%
}\right]  . \label{eq:stensor}%
\end{equation}

The generic spherically symmetric metric can be written as%
\begin{equation}
ds^{2}=-N\left(  r\right)  e^{-2\delta\left(  r\right)  }dt^{2}+\frac{dr^{2}%
}{N\left(  r\right)  }+r^{2}\left(  d\theta^{2}+\sin^{2}\theta d\varphi
^{2}\right)  , \label{eq:metric}%
\end{equation}
where we introduce the Misner-Sharp mass function $m\left(  r\right)  $ as in
$N\left(  r\right)  =1-2m\left(  r\right)  /r$. Due to the spherical symmetry,
the electromagnetic field and the scalar field are given by $A_{\mu}dx^{\mu
}=V\left(  r\right)  dt$ and $\phi=\phi\left(  r\right)  $, respectively. With
this ansatz, the equations of motion $\left(  \ref{eq:eom}\right)  $ then
reduce to%
\begin{gather}
N^{\prime}\left(  r\right)  -\frac{1}{r}\left[  1-N\left(  r\right)  \right]
+r\phi^{\prime}\left(  r\right)  ^{2}N\left(  r\right)  =\frac{2}{aQ^{2}%
}\left[  r^{2}f\left(  \phi\right)  -\sqrt{aQ^{2}+f^{2}\left(  \phi\right)
r^{4}}\right]  ,\nonumber\\
\left[  r^{2}N\left(  r\right)  \phi^{\prime}\left(  r\right)  \right]
^{\prime}+r^{3}\phi^{\prime}\left(  r\right)  ^{3}N\left(  r\right)
=-\frac{Q^{2}r^{2}\dot{f}\left(  \phi\right)  }{aQ^{2}+f^{2}\left(
\phi\right)  r^{4}+f\left(  \phi\right)  r^{2}\sqrt{aQ^{2}+f^{2}\left(
\phi\right)  r^{4}}},\nonumber\\
\delta^{\prime}\left(  r\right)  =-r\phi^{\prime}\left(  r\right)
^{2}\label{eq:eomn}\\
V^{\prime}\left(  r\right)  =-\frac{e^{-\delta\left(  r\right)  }Q}%
{\sqrt{aQ^{2}+f^{2}\left(  \phi\right)  r^{4}}},\nonumber
\end{gather}
where primes denote derivatives with respect to the radial coordinate $r$, and
$Q$ is a constant that can be interpreted as the electric charge.

To find asymptotically flat black hole solutions of the non-linear ordinary
differential equations $\left(  \ref{eq:eomn}\right)  $, one needs to impose
regular boundary conditions at the event horizon and spatial infinity. The
regularity of the solutions across the event horizon at $r=r_{+}$ gives that
the solutions can be approximated by a power series expansion in $r-r_{+}$,
\begin{align}
m\left(  r\right)   &  =\frac{r_{+}}{2}+\left(  r-r_{+}\right)  m_{1}%
+\cdots\text{, }\delta\left(  r\right)  =\delta_{0}+\left(  r-r_{+}\right)
\delta_{1}+\cdots\text{,}\nonumber\\
\phi\left(  r\right)   &  =\phi_{0}+\left(  r-r_{+}\right)  \phi_{1}%
+\cdots\text{, }V\left(  r\right)  =\left(  r-r_{+}\right)  v_{1}%
+\cdots\label{eq:asysmrh}%
\end{align}
where%
\begin{align}
m_{1}  &  =\frac{2\sqrt{aQ^{2}+f^{2}(\phi_{0})r_{+}^{4}}-2f(\phi_{0})r_{+}%
^{2}}{2a},\nonumber\\
\delta_{1}  &  =\frac{Q^{4}\dot{f}^{2}\left(  \phi_{0}\right)  r_{+}%
^{3}\left\{  4\left[  a+2f(\phi_{0})r_{+}^{2}\right]  \left[  \sqrt
{aQ^{2}+f^{2}(\phi_{0})r_{+}^{4}}+f(\phi_{0})r_{+}^{2}\right]  +a\left(
a+4Q^{2}\right)  \right\}  }{\left[  a+4f(\phi_{0})r_{+}^{2}-4Q^{2}\right]
^{2}\left\{  r_{+}^{2}f(\phi_{0})\left[  \sqrt{aQ^{2}+f^{2}(\phi_{0})r_{+}%
^{4}}+f(\phi_{0})r_{+}^{2}\right]  +aQ^{2}\right\}  ^{2}},\nonumber\\
\phi_{1}  &  =-\frac{r_{+}\dot{f}\left(  \phi_{0}\right)  \left[  \sqrt
{aQ^{2}+f^{2}(\phi_{0})r_{+}^{4}}-r_{+}^{2}f(\phi_{0})+2Q^{2}\right]
}{\left[  a+4r_{+}^{2}f(\phi_{0})-4Q^{2}\right]  \sqrt{aQ^{2}+f^{2}(\phi
_{0})r_{+}^{4}}},\\
v_{1}  &  =-\frac{e^{-\delta_{0}}Q}{\sqrt{aQ^{2}+f^{2}(\phi_{0})r_{+}^{4}}%
}.\nonumber
\end{align}
The two parameters, $\phi_{0}$ and $\delta_{0}$, determine the expansion
coefficients and hence the solutions in the vicinity of the horizon. The
Hawking temperature $T_{H}$ and the horizon area $A_{H}$ are given by
\begin{equation}
T_{H}=\frac{N^{\prime}\left(  r_{+}\right)  e^{-\delta\left(  r_{+}\right)  }%
}{4\pi}\text{ and }A_{H}=4\pi r_{+}^{2}, \label{eq:TandA}%
\end{equation}
respectively. At spatial infinity, the asymptotic expansion of the solutions
takes the form,
\begin{equation}
m\left(  r\right)  =M-\frac{Q^{2}+Q_{s}^{2}}{2r}+\cdots\text{, }\delta\left(
r\right)  =\frac{Q_{s}^{2}}{2r^{2}}+\cdots\text{, }\phi\left(  r\right)
=\frac{Q_{s}}{r}+\frac{Q_{s}M}{r^{2}}+\cdots\text{, }V\left(  r\right)
=\Phi+\frac{Q}{r}\text{,} \label{eq:asysminf}%
\end{equation}
where we assume $f\left(  0\right)  =1$. Here $M$ is the the ADM mass, $\Phi$
is the electrostatic potential measured at infinity, and $Q_{s}$ is the scalar
charge. With the asymptotic behavior of the solutions at $r=r_{+}$ and
$r=\infty$, we can use a standard shooting method to solve eqn. $\left(
\ref{eq:eomn}\right)  $ for a family of black hole solutions. Note that the
solutions and the associated physical quantities scale as%
\begin{equation}
r\rightarrow\lambda r\text{, }\phi\rightarrow\phi\text{, }m\rightarrow\lambda
m\text{, }V\rightarrow V\text{, }\delta\rightarrow\delta\text{, }%
Q\rightarrow\lambda Q\text{, }M\rightarrow\lambda M\text{, }a\rightarrow
\lambda^{2}a\text{,} \label{eq:ss}%
\end{equation}
where $\lambda$ is a constant. For later use, we then introduce some reduced
quantities,%
\begin{equation}
q=\frac{Q}{M}\text{, }\tilde{a}=\frac{a}{Q^{2}}\text{, }a_{H}=\frac{A_{H}%
}{16\pi M^{2}}\text{, }t_{H}=8\pi MT_{H},
\end{equation}
which are dimensionless and invariant under the scaling symmetry $\left(
\ref{eq:ss}\right)  $.

The Smarr relation \cite{Smarr:1972kt} relates the black hole mass to other
physical quantities, and can be used to test the accuracy of numerical black
hole solutions. For a manifold $\mathcal{M}$ endowed with the time-like
Killing vector $K^{\mu}$, we consider a hypersurface $\Sigma$ with the
boundary $\partial\Sigma$. Due to Gauss's law and Einstein's equations,
integrating the identity for Killing vectors $\nabla_{\mu}\left(  \nabla_{\nu
}K^{\mu}\right)  =K^{\mu}R_{\mu\nu}$ over the hypersurface $\Sigma$ yields%
\begin{equation}
\int_{\partial\Sigma}dS_{\mu\nu}\nabla^{\mu}K^{\nu}=\frac{1}{2}\int_{\Sigma
}dS_{\mu}K_{\nu}\left(  \mathcal{T}^{\mu\nu}-\frac{1}{2}\mathcal{T}g^{\mu\nu
}\right)  , \label{eq:Kintegral}%
\end{equation}
where $dS_{\mu\nu}$ is the surface element on $\partial\Sigma$, $dS_{\mu}$
denotes the volume element on $\Sigma$, and $\mathcal{T}^{\mu\nu}$ is the
stress-energy tensor. For the metric $\left(  \ref{eq:metric}\right)  $, the
Killing vector $K^{\mu}=\left(  1,0,0,0\right)  $, and we choose the
hypersurface of constant time $t$ bounded by the horizon and spatial infinity
to be $\Sigma$, such that the boundary $\partial\Sigma$ consists of $r=r_{+}$
and $r=+\infty$. Using eqns. $\left(  \ref{eq:stensor}\right)  $ and $\left(
\ref{eq:eomn}\right)  $, we find that the Smarr relation is given by%
\begin{equation}
M=2\pi r_{+}^{2}T_{H}+Q\Phi+\int_{r_{+}}^{\infty}\frac{Q^{2}e^{-\delta\left(
r\right)  }}{\sqrt{\tilde{a}Q^{4}+f^{2}\left(  \phi\right)  r^{4}}}\left[
\frac{2f\left(  \phi\right)  r^{2}}{\sqrt{\tilde{a}Q^{4}+f^{2}\left(
\phi\right)  r^{4}}+f\left(  \phi\right)  r^{2}}-1\right]  dr,
\label{eq:smarrR}%
\end{equation}
where the last term vanishes for the EMS model with $a=0$.

To study perturbative stability of black hole solutions, one can consider
spherically symmetric and time-dependent linear perturbations around the black
hole $\left(  \ref{eq:metric}\right)  $.\ The metric ansatz including the
perturbations can be written as \cite{Fernandes:2019rez}%
\begin{equation}
ds^{2}=-\tilde{N}\left(  r,t\right)  e^{-2\tilde{\delta}\left(  r,t\right)
}dt^{2}+\frac{dr^{2}}{\tilde{N}\left(  r,t\right)  }+r^{2}\left(  d\theta
^{2}+\sin^{2}\theta d\varphi^{2}\right)  ,
\end{equation}
where%
\begin{equation}
\tilde{N}\left(  r,t\right)  =N\left(  r\right)  +\epsilon\tilde{N}_{1}\left(
r\right)  e^{-i\Omega t}\text{ and }\tilde{\delta}\left(  r,t\right)
=\delta\left(  r\right)  +\epsilon\tilde{\delta}_{1}\left(  r\right)
e^{-i\Omega t}\text{.} \label{eq:NdeltaP}%
\end{equation}
The time dependence of the perturbations is assumed to be Fourier modes with
frequency $\Omega$. Similarly for the scalar and BI fields, the ansatz is
given by%
\begin{equation}
\tilde{\phi}\left(  r,t\right)  =\phi\left(  r\right)  +\epsilon\tilde{\phi
}_{1}\left(  r\right)  e^{-i\Omega t}\text{ and }\tilde{V}\left(  r,t\right)
=V\left(  r\right)  +\epsilon\tilde{V}_{1}\left(  r\right)  e^{-i\Omega
t}\text{,} \label{eq:phiVP}%
\end{equation}
respectively. Plugging the ansatzes $\left(  \ref{eq:NdeltaP}\right)  $ and
$\left(  \ref{eq:phiVP}\right)  $ into the equations of motion $\left(
\ref{eq:eom}\right)  $, we find that the gravity equations lead to%
\begin{equation}
\tilde{\delta}_{1}\left(  r\right)  =-2r\phi^{\prime}\left(  r\right)
\tilde{\phi}_{1}^{\prime}\left(  r\right)  \text{, }\tilde{N}_{1}\left(
r\right)  =-2rN\left(  r\right)  \phi^{\prime}\left(  r\right)  \tilde{\phi
}_{1}\left(  r\right)  , \label{eq:NdeltaEom}%
\end{equation}
and the BI field equation gives
\begin{equation}
\tilde{V}_{1}^{\prime}\left(  r\right)  =-V^{\prime}(r)\left\{  \tilde{\phi
}_{1}\left(  r\right)  \frac{\dot{f}\left(  \phi\right)  }{f(\phi)}\left[
1-ae^{2\delta(r)}V^{\prime}(r)^{2}\right]  +\tilde{\delta}_{1}\left(
r\right)  f(\phi)\right\}  . \label{eq:phiVEom}%
\end{equation}
With the help of eqns. $\left(  \ref{eq:eomn}\right)  $, $\left(
\ref{eq:NdeltaEom}\right)  $ and $\left(  \ref{eq:phiVEom}\right)  $, we can
express the perturbation equation for $\tilde{\phi}_{1}\left(  r\right)  $ in
the familiar Schr\"{o}dinger-equation form,%
\begin{equation}
-\frac{d^{2}\Phi}{dr^{\ast2}}+U_{\Omega}\Phi=\Omega^{2}\Phi,
\end{equation}
where $\Phi\equiv r\tilde{\phi}_{1}\left(  r\right)  $, and $r^{\ast}$ is the
tortoise coordinate defined by $dr^{\ast}/dr=e^{\delta\left(  r\right)
}N^{-1}\left(  r\right)  $. The effective potential $U_{\Omega}$ is%
\begin{gather}
U_{\Omega}=\frac{N(r)e^{-2\delta(r)}}{r^{2}}\left\{  1-N(r)-2r^{2}\phi
^{\prime}(r)^{2}+\frac{Q^{2}}{\sqrt{\tilde{a}Q^{4}+f^{2}\left(  \phi\right)
r^{4}}}\left[  \frac{\dot{f}^{2}\left(  \phi\right)  r^{4}}{\tilde{a}%
Q^{4}+f^{2}\left(  \phi\right)  r^{4}}-2+4r^{2}\phi^{\prime}(r)^{2}\right]
\right. \nonumber\\
\left.  +\frac{r^{2}}{a}\left(  \frac{f(\phi)r^{2}}{\sqrt{\tilde{a}Q^{4}%
+f^{2}\left(  \phi\right)  r^{4}}}-1\right)  \left[  4r\phi^{\prime}(r)\dot
{f}\left(  \phi\right)  +2f(\phi)\left(  2r^{2}\phi^{\prime}(r)^{2}-1\right)
+\ddot{f}\left(  \phi\right)  \right]  \right\}  , \label{eq:Ueff}%
\end{gather}
which can be shown to vanish at the event horizon and spatial infinity via the
asymptotic expansions $\left(  \ref{eq:asysmrh}\right)  $ and $\left(
\ref{eq:asysminf}\right)  $. An unstable mode of the EBIS model that has
$\Omega^{2}<0$ would then correspond to a bound state of $U_{\Omega}$. From
quantum mechanics, it follows that if the bound state exists, $\Omega^{2}$
must exceed the minimum value of $U_{\Omega}$. Therefore, when $U_{\Omega}$ is
always positive for $r\in\left(  r_{+},+\infty\right)  $, no bound states
exist, and hence the black hole solutions are stable against spherically
symmetric perturbations. However, note that the appearance of a negative
region in $U_{\Omega}$ does not necessarily mean the presence of radial
instability \cite{Zou:2019bpt}. In this case, one may need other methods, such
as the S-deformation method \cite{Kimura:2017uor}, to investigate the
stability of the solutions.

\section{Born-Infeld Black Hole}

\label{Sec:BIBH}

\begin{figure}[tb]
\begin{center}
\includegraphics[width=0.48\textwidth]{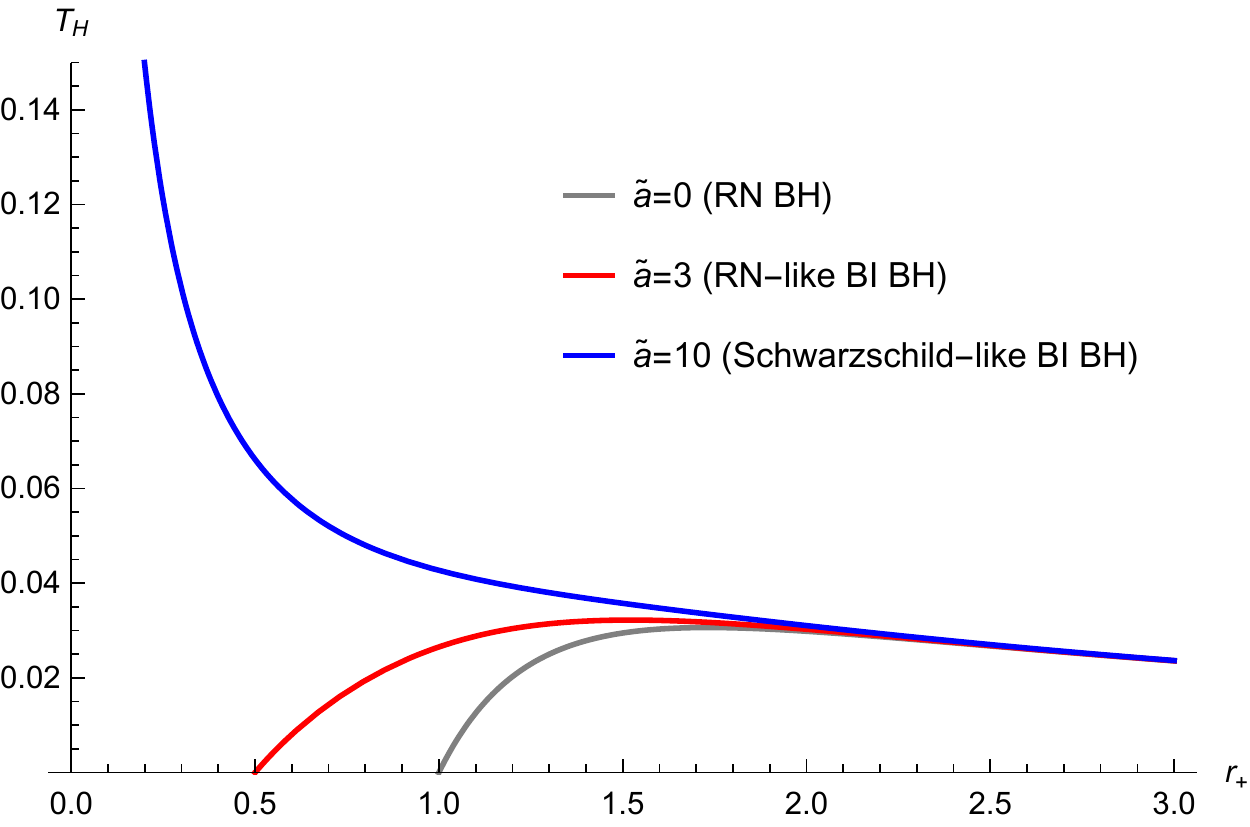}
\includegraphics[width=0.48\textwidth]{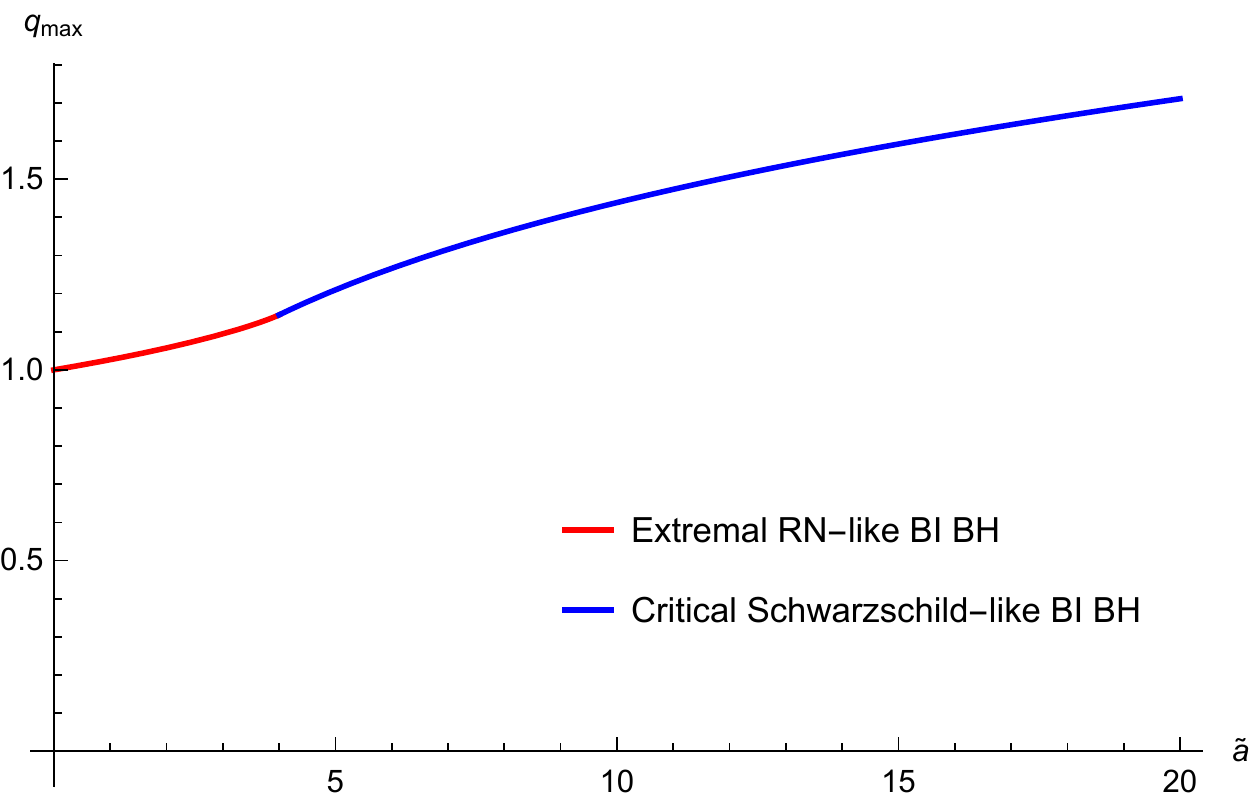}
\end{center}
\caption{{\footnotesize \textbf{Left}: Plot of the Hawking temperature $T_{H}$
against the horizon radius $r_{+}$ for a RN black hole with $\tilde{a}=0$, a
RN-like BI black hole with $\tilde{a}=3$ and a Schwarzschild-like BI black
hole with $\tilde{a}=10$. Here we take $Q=1$. For $\tilde{a}=0$ and $\tilde
{a}=3$, the black holes possess an extremal limit, where $T_{H}=0$. \ For
$\tilde{a}=10$, the black hole temperature behaves like that of a
Schwarzschild black hole, in that it approaches infinity as $r_{+}$ goes to
zero. \textbf{Right}: Plot of the maximum value of the mass-to-charge ratio
$q_{\max}$ against $\tilde{a}$. Extremal RN-like BI black holes have the
largest mass-to-charge ratio, while $q_{\max}$ occurs at critical
Schwarzschild-like BI black holes, which have $r_{+}=0$ while $Q$ remains
finite.}}%
\label{fig:BIBH}%
\end{figure}

In this paper, we study spontaneous scalarization of the EBIS model, which
requires a scalar-free solution. The existence of such a solution then leads
to $\dot{f}\left(  0\right)  =0$, which is obtained via the scalar equation in
eqn. $\left(  \ref{eq:eom}\right)  $. In this section, we consider the
scalar-free solution to eqn. $\left(  \ref{eq:eomn}\right)  $ with $\dot
{f}\left(  0\right)  =0$. When the scalar field $\phi=0$, the static
spherically symmetric black hole solution was first derived in
\cite{Dey:2004yt,Cai:2004eh},%
\begin{gather}
N(r)=1-\frac{2M}{r}-\frac{2Q^{2}}{3\sqrt{r^{4}+\tilde{a}Q^{4}}+3r^{2}}%
+\frac{4Q^{2}}{3r^{2}}\text{ }_{2}F_{1}\left(  \frac{1}{4},\frac{1}{2}%
,\frac{5}{4};-\frac{\tilde{a}Q^{4}}{r^{4}}\right)  \text{,}\nonumber\\
\delta\left(  r\right)  =0\text{ and }V^{\prime}\left(  r\right)  =-\frac
{Q}{\sqrt{r^{4}+\tilde{a}Q^{4}}}, \label{eq:BIBH}%
\end{gather}
where $M$ and $Q$ are the mass and charge of the black hole, respectively. The
Hawking temperature of the BI black hole can be calculated from eqn. $\left(
\ref{eq:TandA}\right)  $:%

\begin{equation}
T_{H}=\frac{1}{4\pi r_{+}}\left(  1-\frac{2Q^{2}}{r_{+}^{2}+\sqrt{r_{+}%
^{4}+\tilde{a}Q^{4}}}\right)  , \label{eq:HTBIBH}%
\end{equation}
where $r_{+}$ is the horizon radius. It showed in \cite{Tao:2017fsy} that
there are two types of BI black holes depending on the minimum value of
$T_{H}$:

\begin{itemize}
\item RN type: $\tilde{a}\leq4$. This type of BI black holes can have extremal
black hole solutions like RN black holes. For $T_{H}=0$, the horizon radius of
the extremal BI black hole is $r_{e}\equiv Q\sqrt{1-\tilde{a}/4}$. We plot
$T_{H}$ versus $r_{+}$ for a RN-like BI black hole with $\tilde{a}=3$ in the
left panel of FIG. \ref{fig:BIBH}, where we have $Q=1$, and the black hole
becomes extremal at $r_{e}=1/2$. Similar to a RN black hole, the
mass-to-charge ratio of a RN-like BI black hole reaches the maximum when the
black hole is extremal. The right panel of FIG. \ref{fig:BIBH} shows the
maximum mass-to-charge ratio $q_{\max}$ versus $\tilde{a}$ for RN-like BI
black holes, which is represented by a red line.

\item Schwarzschild-like type: $\tilde{a}>4$. Like Schwarzschild black holes,
the Hawking temperature $T_{H}$ of this type of BI black holes is always
greater than zero, and diverges as $r_{+}\rightarrow0$. We display $T_{H}$
against $r_{+}$ for a Schwarzschild-like BI black hole with $\tilde{a}=10$ and
$Q=1$ in the left panel of FIG. \ref{fig:BIBH}, which shows that $T_{H}$ is a
monotonically decreasing function of $r_{+}$, and becomes infinite at
$r_{+}=0$. Unlike the RN case, the largest mass-to-charge ratio $q_{\max}$
occurs when $r_{+}=0$ with $Q$ being finite. The BI black holes that have
vanishing $r_{+}$ and finite $Q$ (or $M$) are dubbed as\ critical BI black
holes.\ In the right panel of FIG. \ref{fig:BIBH}, $q_{\max}$ versus
$\tilde{a}$ for Schwarzschild-like BI black holes is depicted by a blue line.
\end{itemize}

To study the stability of the BI black hole solution against a scalar
perturbation $\delta\phi$, we linearize the scalar equation in eqn. $\left(
\ref{eq:eom}\right)  $ around the scalar-free solution, which gives
\begin{equation}
\frac{\partial^{\mu}\left(  \sqrt{-g}\partial_{\mu}\delta\phi\right)  }%
{\sqrt{-g}}=\mu_{eff}^{2}\text{ with }\mu_{eff}^{2}=-\ddot{f}\left(  0\right)
\frac{1-r^{2}/\sqrt{r^{4}+\tilde{a}Q^{4}}}{aQ^{2}}. \label{eq:deltaphi}%
\end{equation}
If $\mu_{eff}^{2}<0$, a tachyonic instability is induced, and a scalarized
black hole solution can bifurcate from the scalar-free BI black hole solution.
Note that $\mu_{eff}^{2}<0$ leads to $\ddot{f}\left(  0\right)  >0$. In the
remainder of the paper, we focus an exponential coupling, $f\left(
\phi\right)  =e^{\alpha\phi^{2}}$ with $\alpha>0$, which satisfies $\dot
{f}\left(  0\right)  =0$ and $\ddot{f}\left(  0\right)  >0$.

\begin{figure}[ptbh]
\begin{center}
\includegraphics[width=0.48\textwidth]{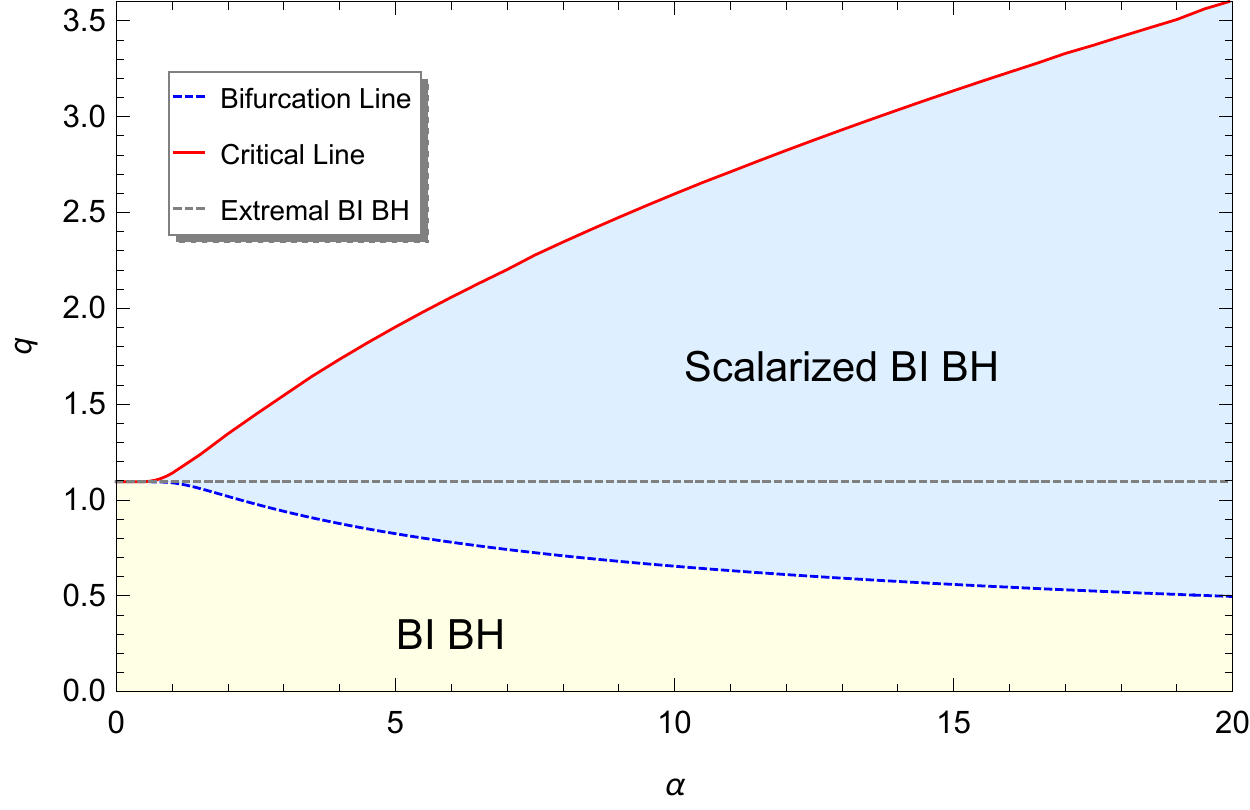}
\includegraphics[width=0.48\textwidth]{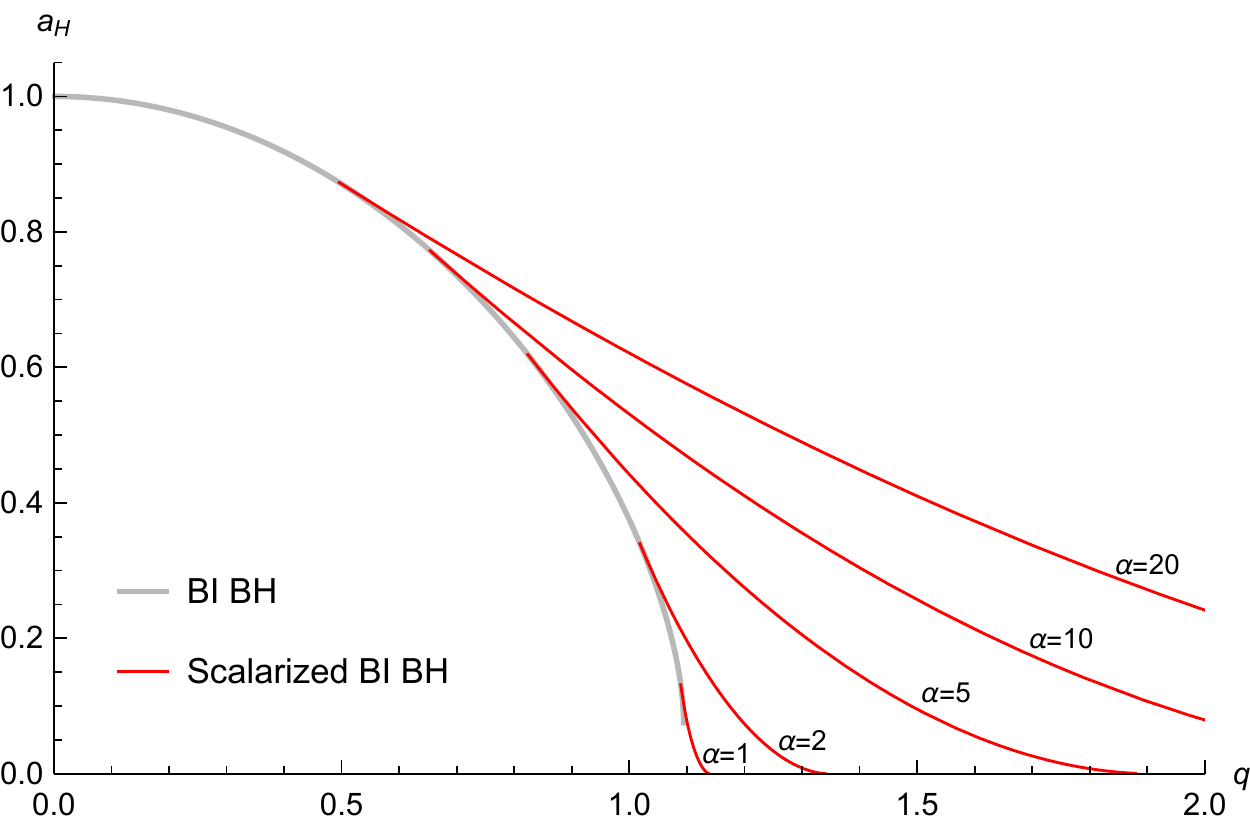}
\includegraphics[width=0.48\textwidth]{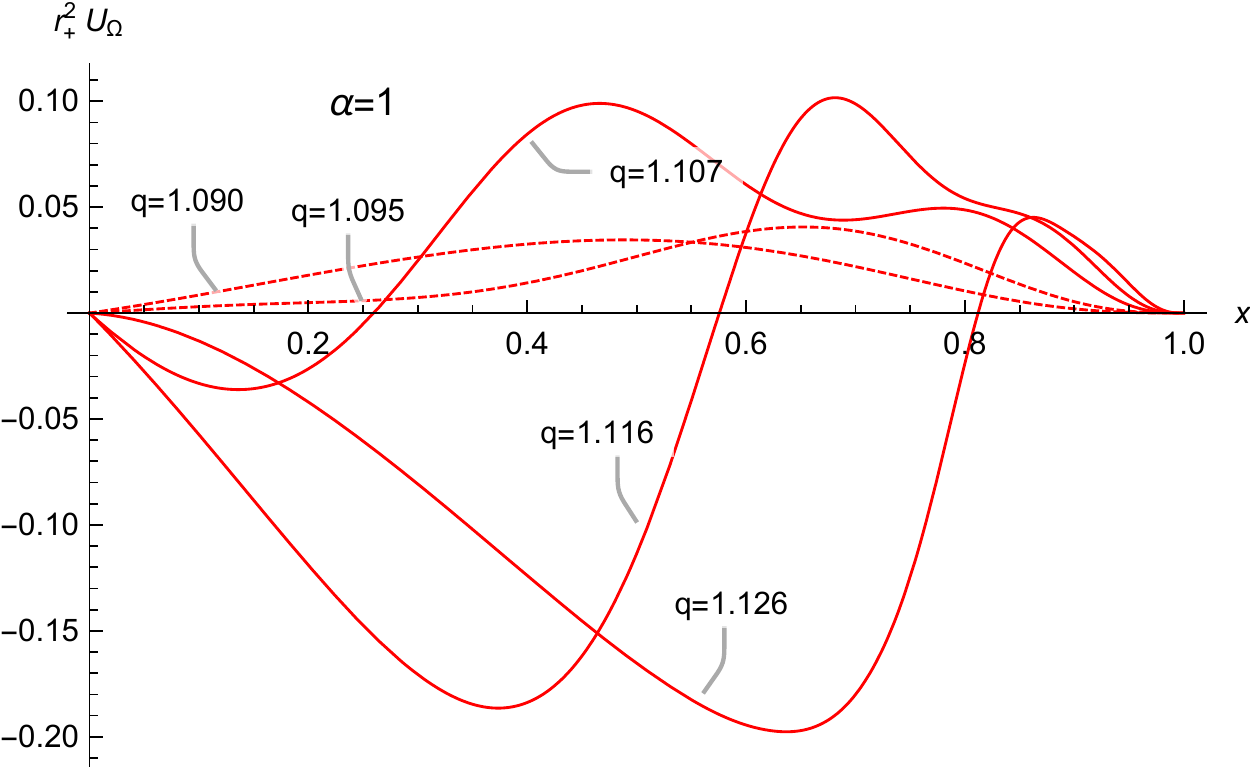}
\includegraphics[width=0.48\textwidth]{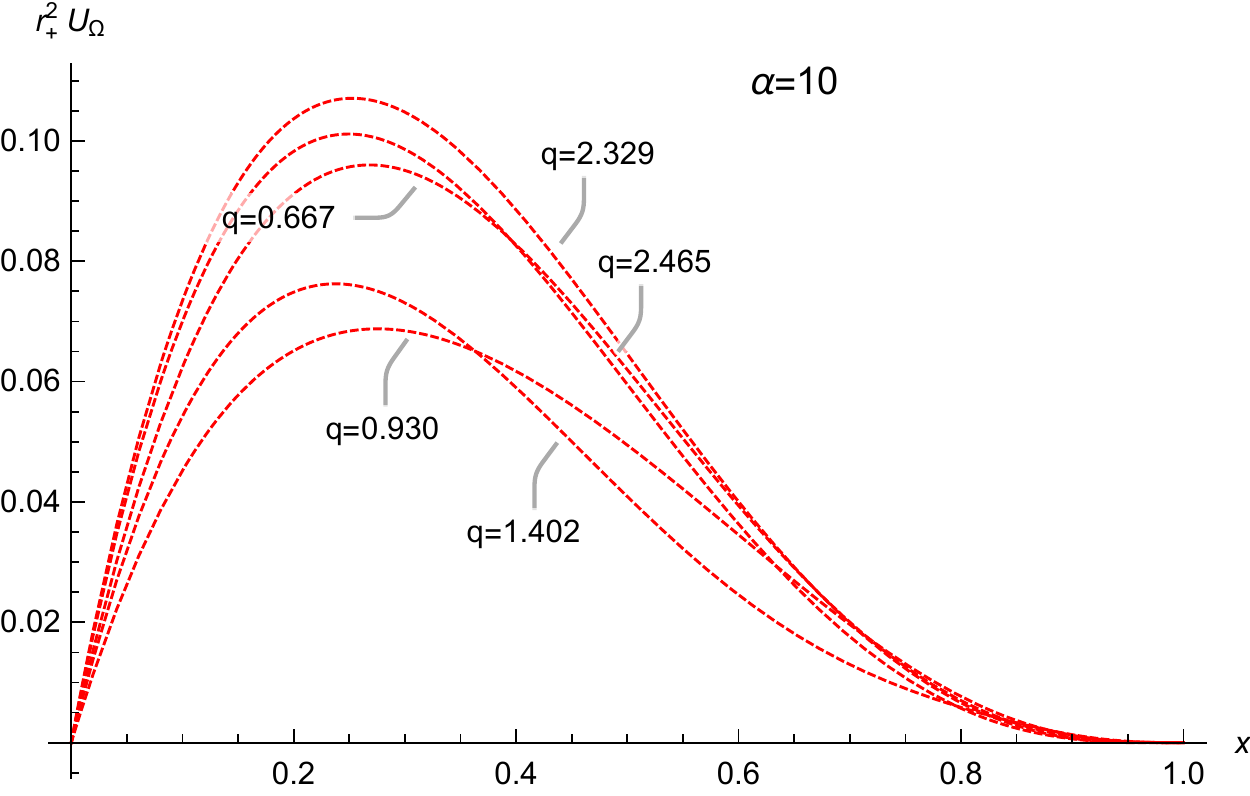}
\end{center}
\caption{{\footnotesize Domain of existence, thermodynamic preference and
effective potentials for scalarized RN-like BI black hole solutions with
$\tilde{a}=3$. \textbf{Upper Left:} Domain of existence in the $\alpha$-$q$
plane, which is displayed by a shaded light blue region and bounded by the
bifurcation and critical lines. The blue dashed line represents the
bifurcation line, where scalarized black holes bifurcate from BI black holes
as zero modes. The red line marks critical configurations of scalarized black
holes, where the horizon area vanishes with the mass remaining finite. In the
light blue region, only one branch of scalarized black hole solutions exists
for a given $\alpha$. \textbf{Upper Right: }Reduced area $a_{H}$ versus
reduced charge $q$ for BI black holes (a gray line) and scalarized BI black
holes with various values of $\alpha$ (red lines). For a given $q$, $a_{H}$ of
the scalarized BI black hole is larger than that of the BI black hole, and
increases with the growth of $\alpha$. The scalarized solutions are always
entropically preferred. \textbf{Lower}: Effective potentials for scalarized BI
black holes with $\alpha=1$ (\textbf{Left}) and $\alpha=10$ (\textbf{Right})
for various values of $q$. Dashed red lines represent positive definite
effective potentials, whereas effective potentials that have negative regions
are exhibited by solid red lines. The $\alpha=10$ scalarized solutions are
stable against radial perturbations, while the stability of the $\alpha=1$
scalarized solutions is inconclusive.}}%
\label{fig:DoEa3}%
\end{figure}

To study the onset of spontaneous scalarization from the scalar-free solution,
we derive the zero mode of BI black holes. First, we express the scalar
perturbation $\delta\phi$ as a spherical harmonics decomposition%
\begin{equation}
\delta\phi=\sum\limits_{l,m}Y_{lm}\left(  \theta,\phi\right)  U_{l}\left(
r\right)  .
\end{equation}
With this decomposition, the scalar equation $\left(  \ref{eq:deltaphi}%
\right)  $ then simplifies to%
\begin{equation}
\frac{\partial_{r}\left[  r^{2}N\left(  r\right)  U_{l}^{\prime}\left(
r\right)  \right]  }{r^{2}}-\left[  \frac{l\left(  l+1\right)  }{r^{2}}%
+\mu_{eff}^{2}\right]  U_{l}\left(  r\right)  =0,
\end{equation}
where $N\left(  r\right)  $ is given by eqn. $\left(  \ref{eq:BIBH}\right)  $.
In addition, $U_{l}\left(  r\right)  $ is regular at $r=r_{+}$ and vanishes at
$r=\infty$. With fixed $\alpha$ and $l$, the boundary conditions of
$U_{l}\left(  r\right)  $ would pick up a set of BI black hole solutions with
different reduced charge $q$. The black hole solutions can be labelled by a
non-negative integer $n$, and $n=0$ is the fundamental mode, whereas $n>0$
corresponds to overtones. In this paper, we focus on the $l=0=n$ mode since it
gives the smallest $q$ of the black hole solutions for a given $\alpha$
\cite{Herdeiro:2018wub}. The reduced charge $q_{\text{exist}}\left(
\alpha\right)  $ of the $l=0=n$ mode compose the bifurcation line in the
$\alpha$-$q$ plane, on which scalarized black hole solutions emerge from the
BI black holes. In the upper left panel of FIG. \ref{fig:DoEa3}, the blue
dashed line represents the bifurcation line of a RN-like BI black hole with
$\tilde{a}=3$, which is quite similar to the RN black hole case
\cite{Herdeiro:2018wub}. The bifurcation line of a Schwarzschild-like BI black
hole with $\tilde{a}=10$ is shown by a blue dashed line (hardly distinguished
from the cyan line for large $\alpha$) in FIG. \ref{fig:DoEa10}. When $\alpha$
is large, the bifurcation lines in the RN- and Schwarzschild-like cases are
almost same. Interestingly for small $\alpha$, the bifurcation line in the
Schwarzschild-like case bears little resemblance to that in the RN-like case.
Specifically, the right panel of FIG. \ref{fig:DoEa10} shows that
Schwarzschild-like BI black holes with $\tilde{a}=10$ and $\alpha\in\left(
2.846,3.139\right)  $ have two $l=0=n$ modes, which marks the onset of two
families of scalarized black hole solutions.

\section{Scalarized Born-Infeld Black Hole}

\label{Sec:SBIBH}

In this section, we present the numerical results, e.g., domains of existence,
thermodynamic preference and effective potentials, for scalarized black hole
solutions, which are dynamically induced from RN-like BI black holes with
$\tilde{a}=3$ and Schwarzschild-like BI black holes with $\tilde{a}=10$. To
solve the non-linear differential equations $\left(  \ref{eq:eomn}\right)  $
for the scalarized black hole solutions, we express the differential equations
$\left(  \ref{eq:eomn}\right)  $ in terms of a new coordinate%
\begin{equation}
x=1-\frac{r_{+}}{r}\text{ with }0\leq x\leq1\text{,}%
\end{equation}
and employ the NDSolve function in Wolfram Mathematica to numerically solve
the equations in the interval $10^{-8}\leq x\leq1$. Solving the differential
equations $\left(  \ref{eq:eomn}\right)  $ numerically is quite standard
except possible stiffness encountered around $x=1$. So we use the
\textquotedblleft StiffnessSwitching\textquotedblright\ method, which by
default switches between the backward differentiation formula and Adams
methods, depending on whether the system being solved is stiff or not. To test
the accuracy of the numerical method, we consider the Smarr relation $\left(
\ref{eq:smarrR}\right)  $, and our numerical results exhibit that the
numerical error can be maintained around the order of $10^{-6}$. In what
follows, we confine ourselves to the simplest case of nodeless, spherically
symmetric black hole solutions and leave general configurations for future work.

\subsection{Scalarized RN-like Born-Infeld Black Hole}

In the upper left panel of FIG. \ref{fig:DoEa3}, we present the domain of
existence for scalarized RN-like BI black holes with $\tilde{a}=3$. For a
given $\alpha$, scalarized solutions emerge from the bifurcation line as zero
modes, and can be continuously induced by increasing $q$ until they reach the
critical line. Our numerical results suggest that, for scalarized solutions on
the critical line, the horizon radius $r_{+}$ vanishes, whereas the mass $M$
and the charge $Q$ remain finite. The domain of existence for scalarized
solutions is bounded by the bifurcation and critical lines, and shows a close
resemblance to that of RN black holes \cite{Herdeiro:2018wub}. For given
$\alpha$ and $q$ in the domain of existence, the numerics show that there
exists a unique set of nodeless scalarized solutions. We plot the reduced area
$a_{H}$ of the scalarized solutions against the reduced charge $q$ for several
values of $\alpha$ in the upper right panel of FIG. \ref{fig:DoEa3}, which
indicates that scalarized solutions are entropically preferred over BI black
hole solutions.

To study stability of scalarized solutions, we depict effective potentials for
scalarized solutions with $\alpha=1$ and $\alpha=10$ in the lower row of FIG.
\ref{fig:DoEa3}. When $\alpha=1$, the scalarized solutions have positive
effective potentials for small enough value of $q$, and thus are free of
radial instabilities. However, as $q$ increases, negative regions in the
effective potentials appear, and hence radial instabilities cannot be
excluded. Moreover, the negative regions of the potentials becomes larger when
moving toward the critical line, i.e., for larger $q$. On the other hand, the
effective potentials of the scalarized solutions with $\alpha=10$ are shown to
be positive, which indicates that the scalarized solutions are radially
stable. Note that solid and dashed colored lines, which represent effective
potentials in the following figures, correspond to potentials with and without
negative regions, respectively.

\subsection{Scalarized Schwarzschild-like Born-Infeld Black Hole}

\begin{figure}[ptbh]
\begin{center}
\includegraphics[width=0.48\textwidth]{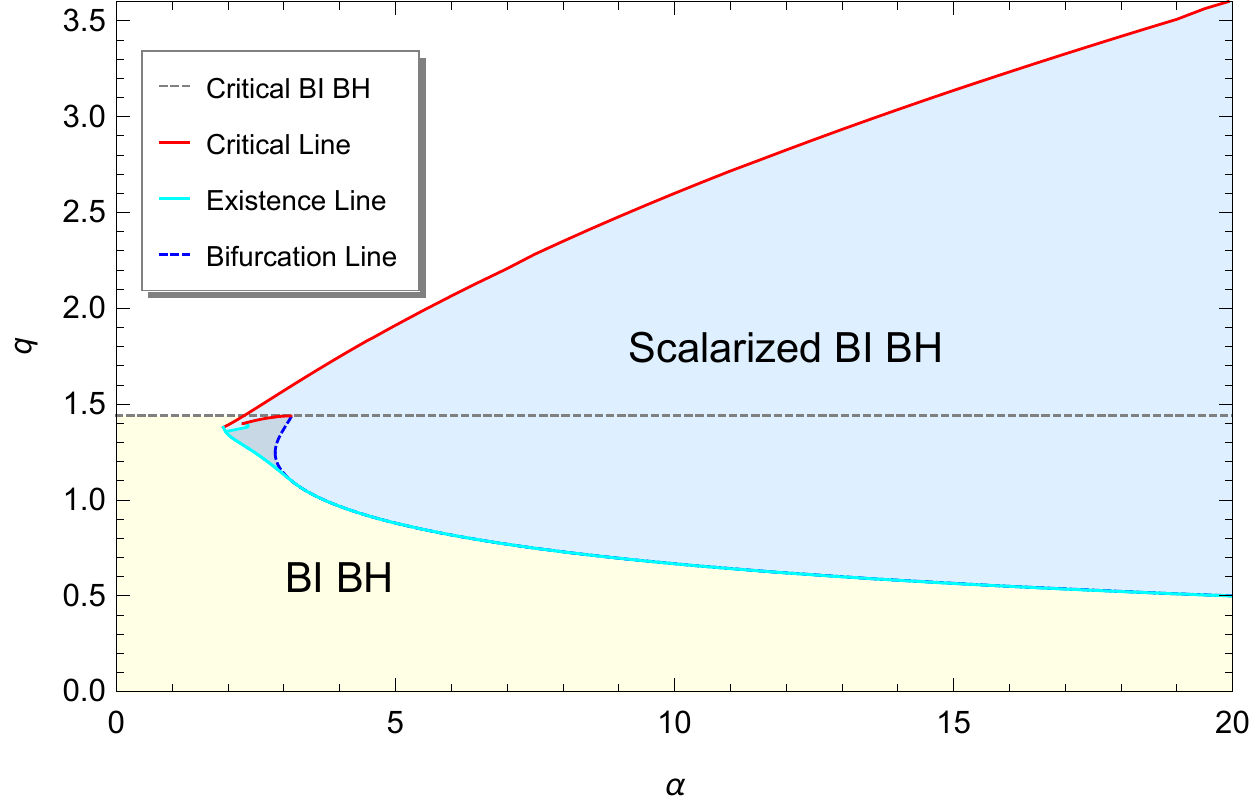}
\includegraphics[width=0.48\textwidth]{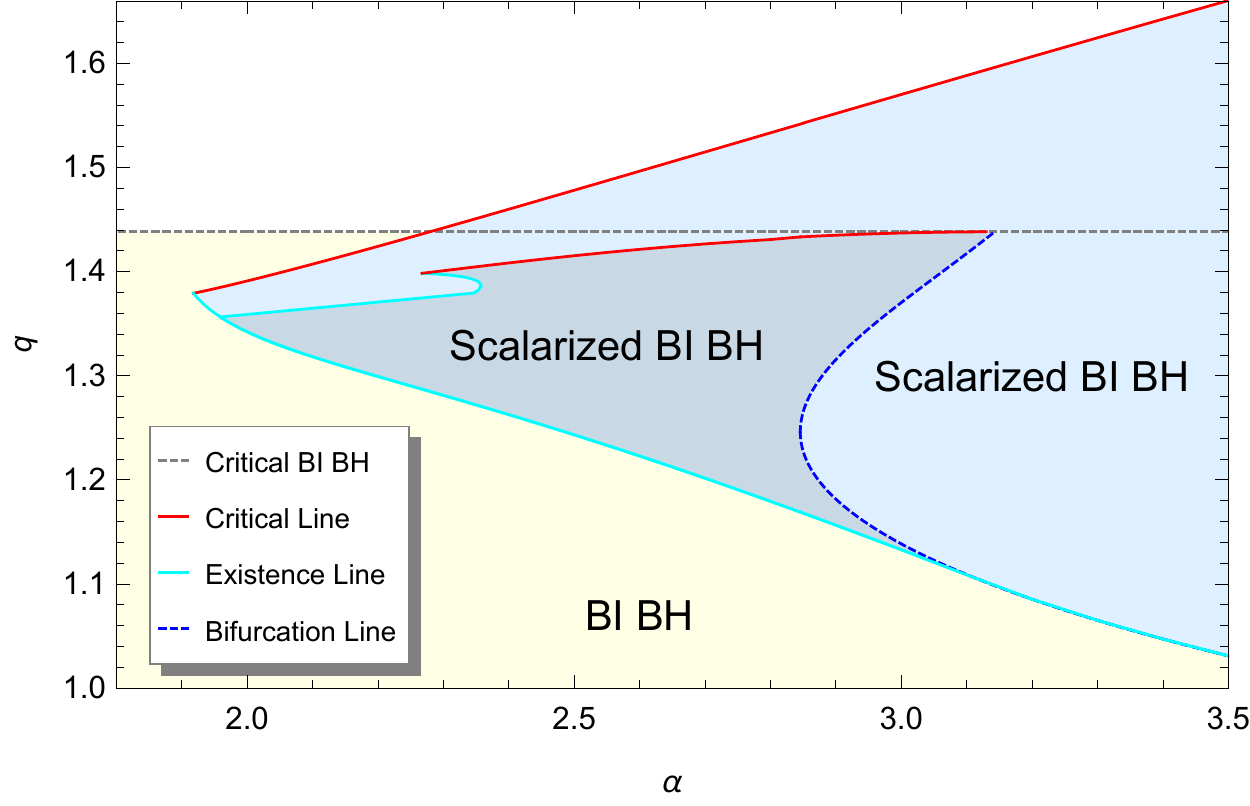}
\end{center}
\caption{{\footnotesize Domain of existence for scalarized Schwarzschild-like
BI black hole solutions with $\tilde{a}=10$. For a given $\alpha$, scalarized
black hole solutions possess two branches in the dark shaded blue region and
one branch in the light one. The domain of existence is delimited by the
critical (red lines) and existence (cyan lines) lines. While scalarized
solutions emerge with vanishing scalar fields on the bifurcation line (a blue
dashed line), scalarized solutions on the existence lines have non-vanishing
scalar field configurations. When $\alpha$ is large, the bifurcation and
existence lines are hardly distinguishable from each other, and the domain of
existence is quite similar to that in the $\tilde{a}=3$ case. The right panel
headlights the small $\alpha$ region, where the domains of existence in the
RN-like and Schwarzschild-like cases are significantly different.}}%
\label{fig:DoEa10}%
\end{figure}

The domain of existence in the $\alpha$-$q$ plane for scalarized
Schwarzschild-like BI black holes with $\tilde{a}=10$ is displayed in FIG.
\ref{fig:DoEa10}, where the right panel highlights small $\alpha$ regime.
Apart from the existence and critical lines, which appear in the RN-like case,
a new type of boundaries of the domain of existence, dubbed as
\textquotedblleft existence line\textquotedblright, is present in the
Schwarzschild-like case. Unlike the bifurcation line, scalarized solutions on
the existence line are not zero modes, and have non-zero finite scalar field
(except $x=1$). The domain of existence for scalarized solutions with
$\tilde{a}=10$ is bounded by the existence, bifurcation and critical lines,
and have more complicated structure than the RN-like case. As $q$ varies,
scalarized solutions in the dark blue regions of FIG. \ref{fig:DoEa10} have
two $\alpha$-constant branches, while these in the light blue regions have
only one $\alpha$-constant branch.

\begin{figure}[ptbh]
\begin{center}
\includegraphics[width=0.48\textwidth]{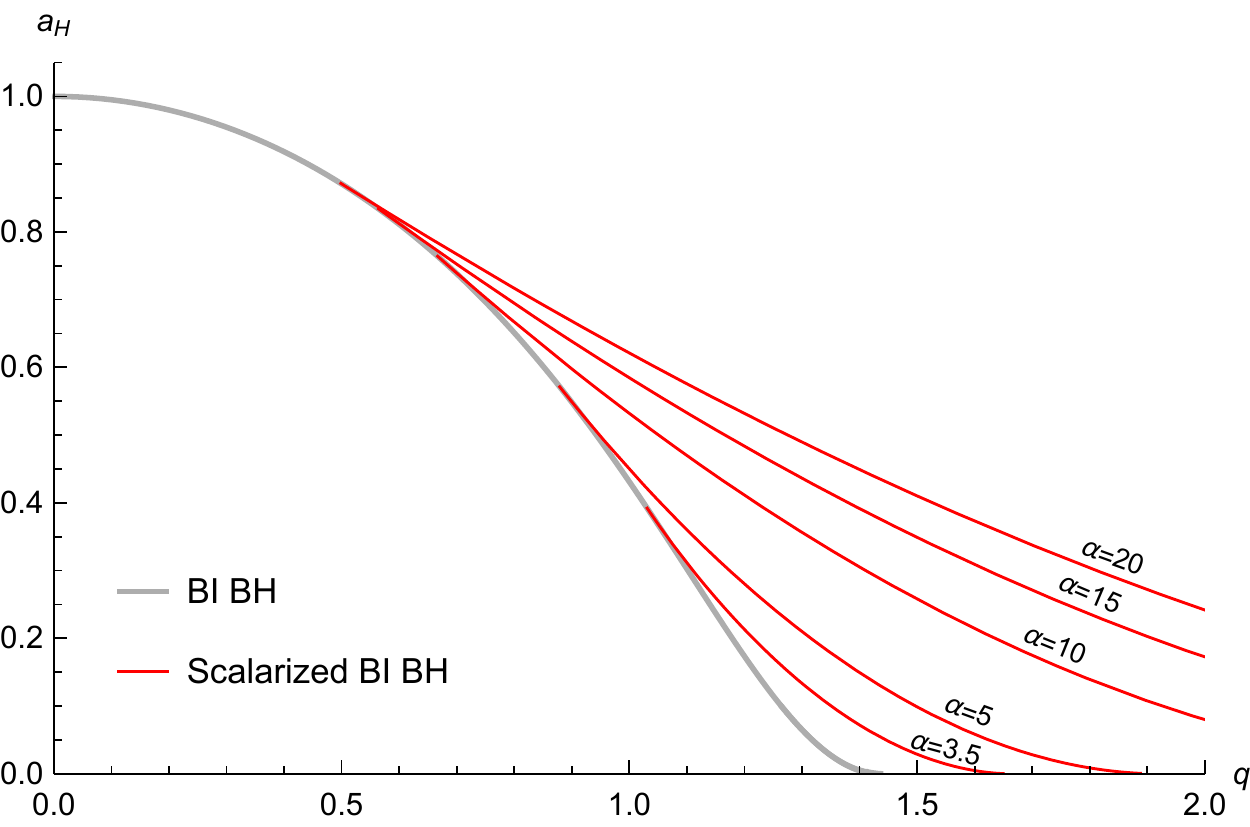}
\includegraphics[width=0.48\textwidth]{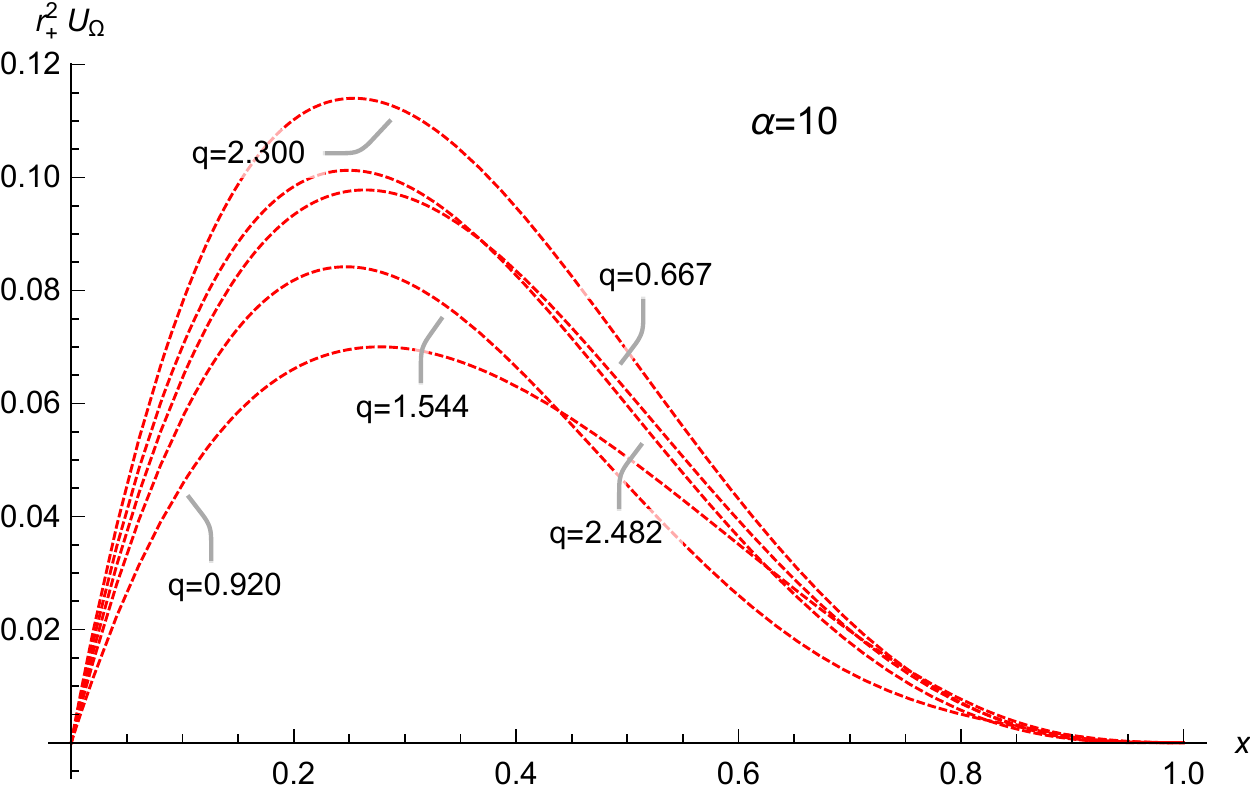}
\end{center}
\caption{{\footnotesize Thermodynamic preference and effective potentials for
Schwarzschild-like BI black hole solutions with $\tilde{a}=10$ in the large
$\alpha$ regime. \textbf{Left:} Reduced area $a_{H}$ versus reduced charge $q$
for BI black holes (a gray line) and scalarized BI black holes with various
values of $\alpha\geq3.5$ (red lines). The scalarized solutions are always
entropically preferred. \textbf{Right}: Effective potentials for scalarized
solutions with $\alpha=10$ for various values of $q$, which are shown to be
positive. The scalarized BI black hole solutions are stable against radial
perturbations.}}%
\label{fig:a10alpha10}%
\end{figure}

When $\alpha$ is large enough, our numerical results barely discriminate
between the existence and bifurcation lines. Practically for a given $a$,
scalarized black holes start from the bifurcation (existence) lines, and form
the single branch of solutions with monotonically increasing $q$ until the
critical line is reached. We plot the reduced area $a_{H}$ against the reduced
charge $q$ for scalarized solutions with several values of $\alpha$ in the
left panel of FIG. \ref{fig:a10alpha10}, which displays that the scalarized
solutions have larger area than BI black holes, and hence are entropically
preferred. Effective potentials of scalarized solutions with $\alpha=10$ are
presented for various values of $q$ in the right panel of FIG.
\ref{fig:a10alpha10}. It exhibits that the effective potentials are positive,
and hence the scalarized solutions are stable against radial perturbations.
Our results suggest that, in large $\alpha$ regime, the behavior of scalarized
RN-like and Schwarzschild-like black hole solutions is quite alike.

\begin{figure}[ptbh]
\begin{center}
\includegraphics[width=0.48\textwidth]{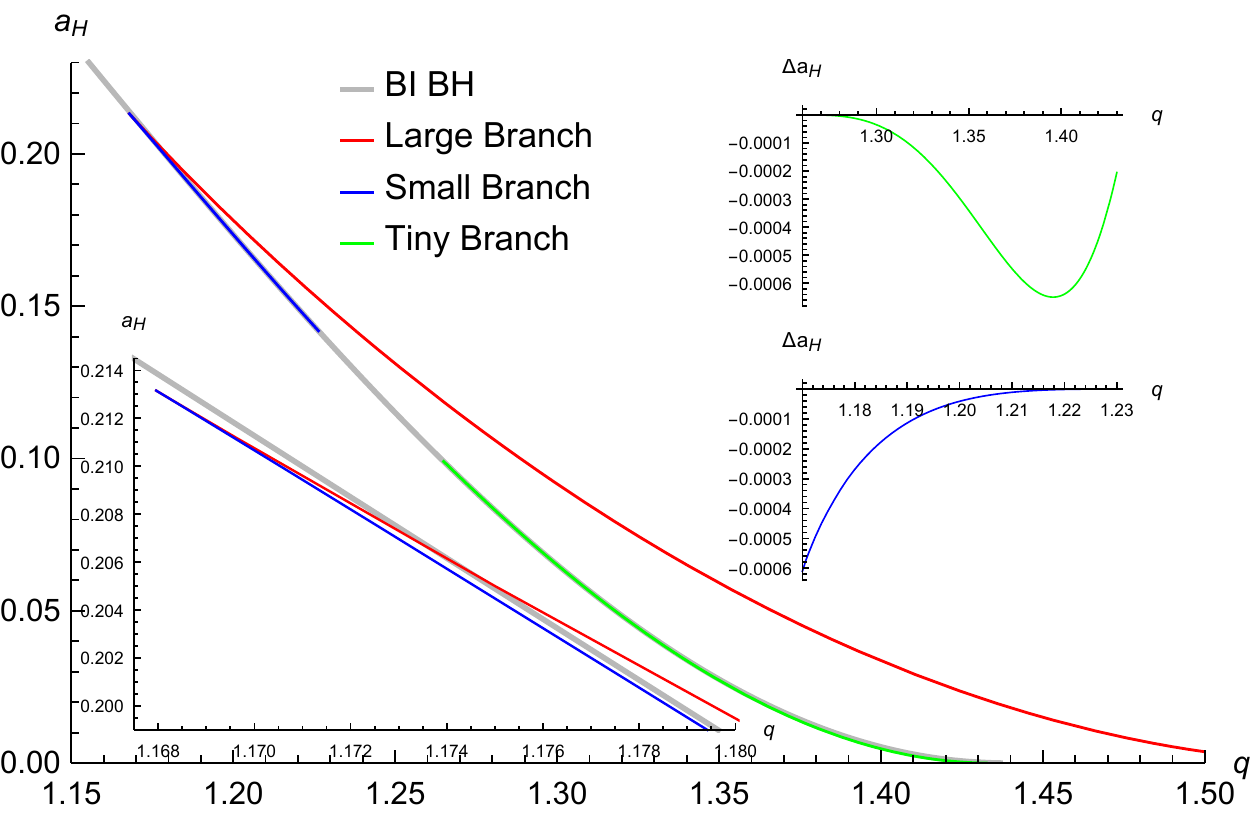}
\includegraphics[width=0.48\textwidth]{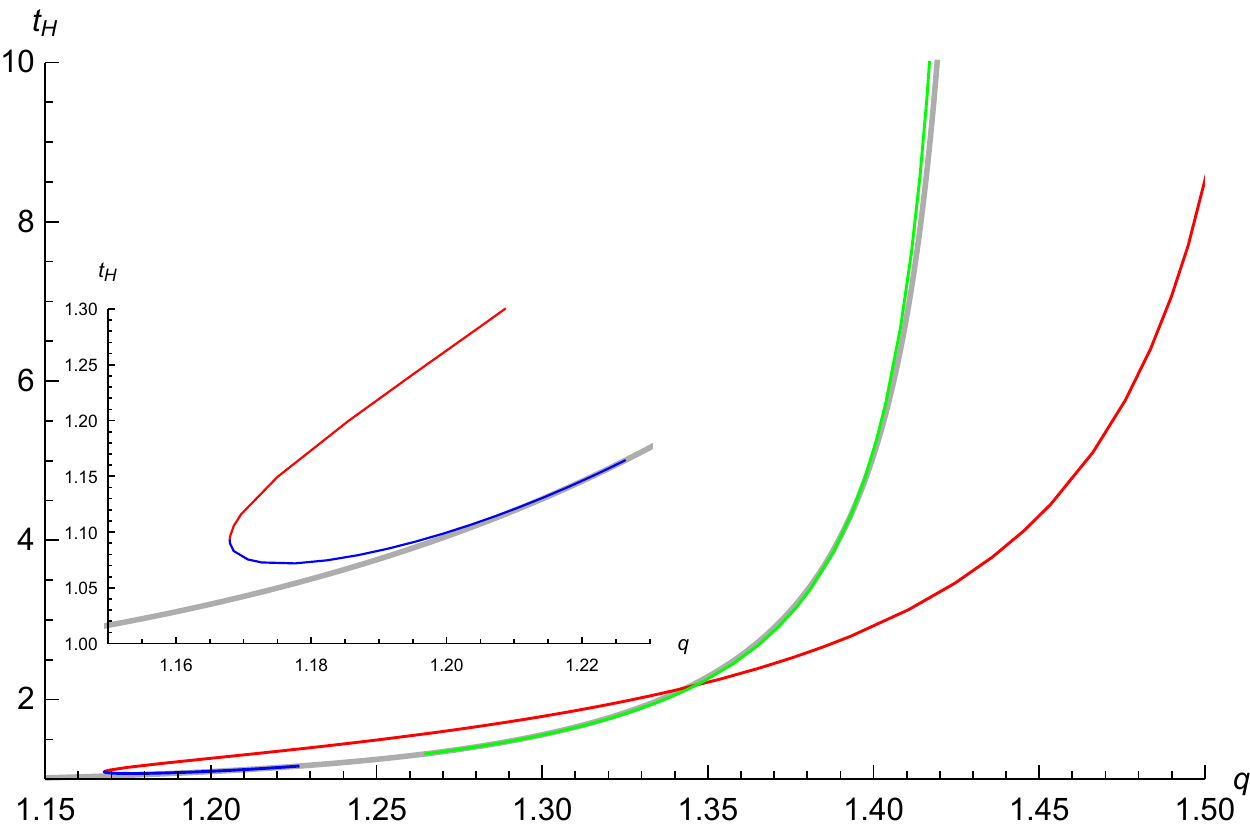}
\includegraphics[width=0.32\textwidth]{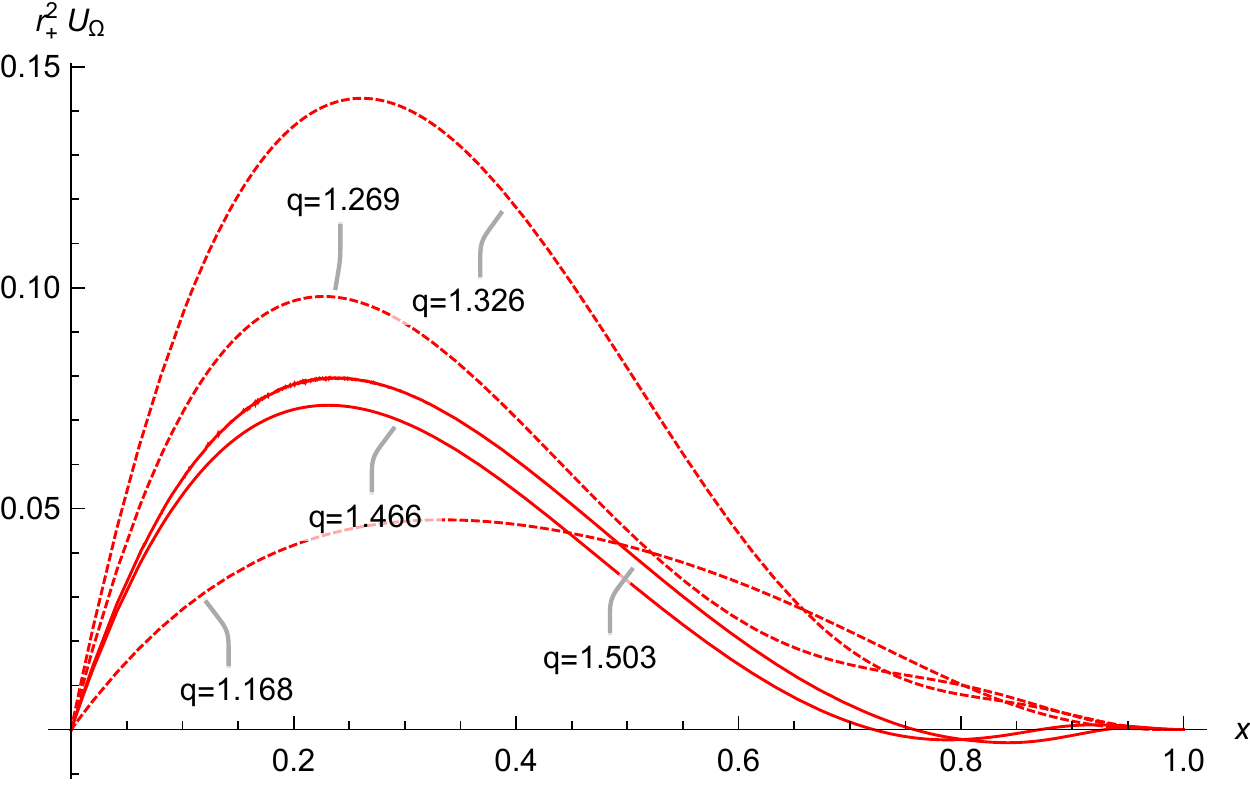}
\includegraphics[width=0.32\textwidth]{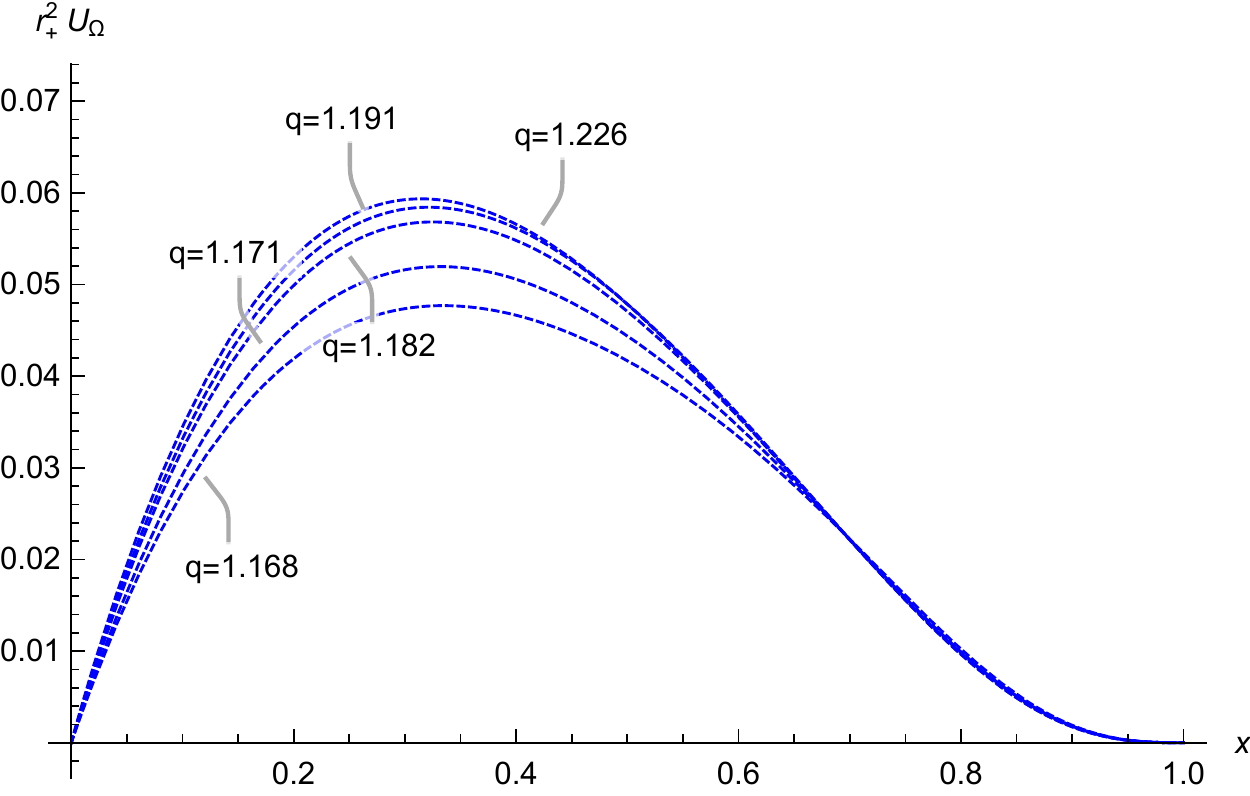}
\includegraphics[width=0.32\textwidth]{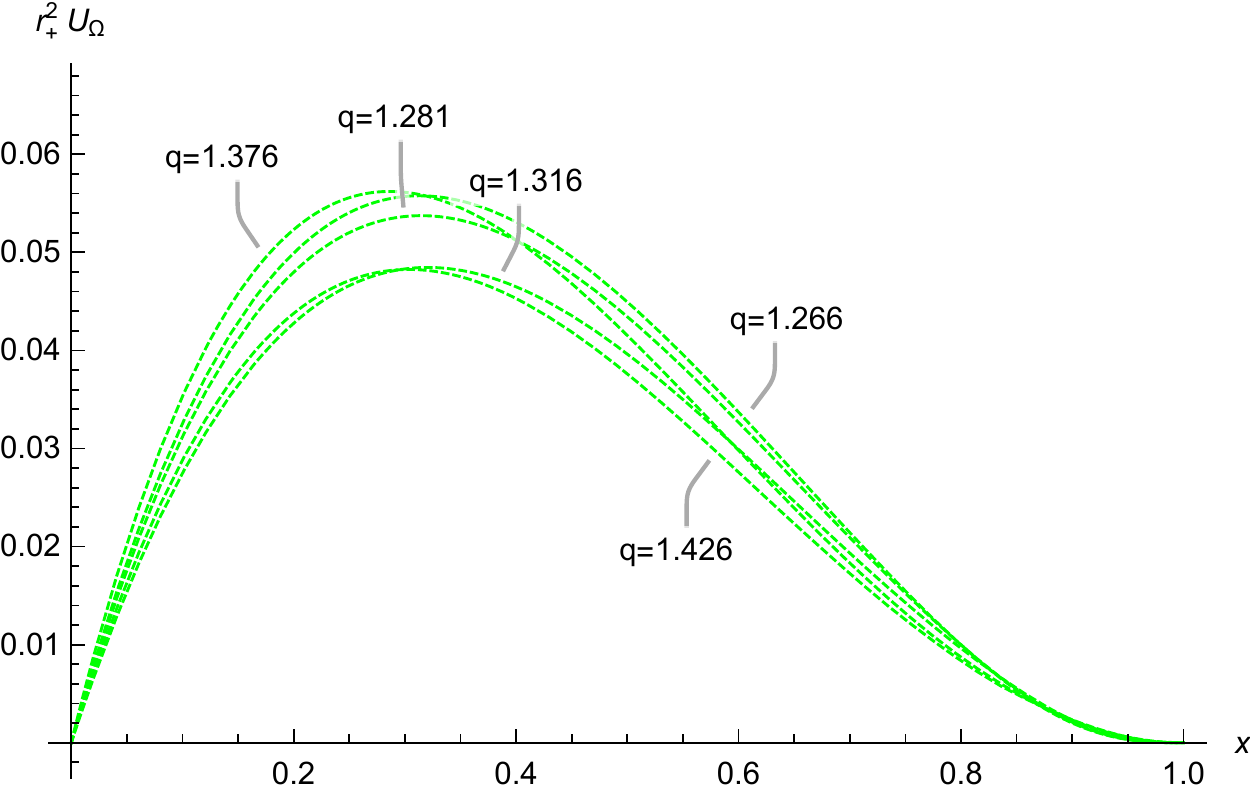}
\end{center}
\caption{{\footnotesize Thermodynamic preference, temperature and effective
potentials for the large (red lines), small (blue lines) and tiny (green
lines) branches of Schwarzschild-like BI black hole solutions with $\tilde
{a}=10$ and $\alpha=2.85$. The large branch coexists with the small and tiny
branches, respectively. \textbf{Upper Left: }Reduced area $a_{H}$ versus
reduced charge $q$ for BI black holes (a gray line) and the three branches of
scalarized BI black holes. The difference of the reduced areas of the
scalarized and BI black holes $\Delta a_{H}\equiv a_{H}^{\text{scalarized BH}%
}-a_{H}^{\text{BI BH}}$ is plotted against $q$ for the small and tiny branches
in the insets on the right of the panel, which show that the small and tiny
branches have smaller area than BI black holes for same $q$. The inset on the
left side displays that most of the large branch is entropically preferred.
\textbf{Upper Right: }Reduced temperature $t_{H}$ versus reduced charge $q$
for BI black holes and the three branches of scalarized BI black holes. For a
given $q$, the large branch is hotter than the small branch, while large
branch is colder than the tiny branch when $q$ is large enough. \textbf{Lower}%
: Effective potentials for the three branches of scalarized BI black holes
with various values of $q$. The small and tiny branches are stable against
radial perturbations. When $q$ is small, the large branch is also stable.
However for large $q$, the stability of the large branch is inconclusive.}}%
\label{fig:a10alpha285}%
\end{figure}

For $2.846\lesssim\alpha\lesssim3.138$, we observe from the right panel of
FIG. \ref{fig:DoEa10} that the bifurcation (blue dashed) line is multi-valued
for a given $\alpha$, i.e., two reduced charges, $q_{\text{bi}}^{\text{up}%
}\left(  \alpha\right)  $ and $q_{\text{bi}}^{\text{low}}\left(
\alpha\right)  $ with $q_{\text{bi}}^{\text{up}}\left(  \alpha\right)
>q_{\text{bi}}^{\text{low}}\left(  \alpha\right)  $, correspond to a same
$\alpha$ on the bifurcation line. The scalarized black holes with
$q_{\text{bi}}^{\text{up}}\left(  \alpha\right)  $ and $q_{\text{bi}%
}^{\text{low}}\left(  \alpha\right)  $ give rise to three branches of
scalarized solutions, namely large branch, small branch and tiny branch, in
the domain of existence. With increasing $q$, the tiny branch emerges from the
scalarized black holes with $q_{\text{bi}}^{\text{up}}\left(  \alpha\right)  $
on the bifurcation line, and ends at a singular configuration with
$q_{\text{cr}}^{\text{low}}\left(  \alpha\right)  $ on the red lower critical
line in FIG. \ref{fig:DoEa10}. The tiny branch then exists for $q_{\text{bi}%
}^{\text{up}}\left(  \alpha\right)  \leq q\leq q_{\text{cr}}^{\text{low}%
}\left(  \alpha\right)  $. The small branch starts from zero modes with
$q_{\text{bi}}^{\text{low}}\left(  \alpha\right)  $ on the bifurcation line
and have a decreasing reduced charge $q$ until a minimum value, $q_{\text{ex}%
}\left(  \alpha\right)  $, is achieved. The curve $q_{\text{ex}}\left(
\alpha\right)  $ is the cyan existence line in FIG. \ref{fig:DoEa10}. On the
existence line, the scalarized solutions bifurcates, and a new\ branch of
solutions, i.e., the large branch, appears. As $q$ increases from
$q_{\text{ex}}\left(  \alpha\right)  $, the large branch reaches out into the
region beyond the existence of BI black holes and terminates at the the upper
critical line with $q_{\text{cr}}^{\text{upp}}\left(  \alpha\right)  $. The
large and small branches exist for $q_{\text{ex}}\left(  \alpha\right)  \leq
q\leq q_{\text{cr}}^{\text{upp}}\left(  \alpha\right)  $ and $q_{\text{ex}%
}\left(  \alpha\right)  \leq q\leq q_{\text{bi}}^{\text{low}}\left(
\alpha\right)  $, respectively. Since $q_{\text{ex}}\left(  \alpha\right)
<q_{\text{bi}}^{\text{low}}\left(  \alpha\right)  $ and $q_{\text{bi}%
}^{\text{up}}\left(  \alpha\right)  <q_{\text{cr}}^{\text{low}}\left(
\alpha\right)  <q_{\text{cr}}^{\text{upp}}\left(  \alpha\right)  $, the large
branch coexists with the small (tiny) branch when $q_{\text{ex}}\left(
\alpha\right)  \leq q\leq q_{\text{bi}}^{\text{low}}\left(  \alpha\right)  $
($q_{\text{bi}}^{\text{up}}\left(  \alpha\right)  \leq q\leq q_{\text{cr}%
}^{\text{low}}\left(  \alpha\right)  $).

To illustrate the properties of the three branches, we plot the reduced area
$a_{H}$ against the reduced charge $q$, the reduced temperature $t_{H}$
against the reduced charge $q$ and effective potentials for the three branches
with $\alpha=2.85$ in FIG. \ref{fig:a10alpha285}. The insets in the upper left
panel of FIG. \ref{fig:a10alpha285} show that the small and tiny branches
always have smaller area (entropy) than the BI black holes with the same $q$.
However except the small part around the existence line, most of the large
branch have larger area and hence is entropically preferred. In addition, the
upper left panel of FIG. \ref{fig:a10alpha285} indicates that the large (tiny)
branch has the largest (smallest) area, which justifies the terms, large,
small and tiny, for the three branches. The reduced temperature $t_{H}$
against the reduced charge $q$ is depicted for the three branches and BI black
holes in the upper right panel of FIG. \ref{fig:a10alpha285}. All the
temperatures monotonically increase as $q$ increases. In the coexistence
region, the large branch is hotter than the small branch, whereas both
branches are hotter than BI black holes. There is a crossing of $t_{H}$ of the
large and tiny branches between $q_{\text{ex}}^{\text{up}}\left(
\alpha\right)  $ and $q_{\text{cr}}^{\text{low}}\left(  \alpha\right)  $. For
large (small) $q$, the large branch is colder (hotter) than the tiny branch
and BI black holes. The effective potentials for the three branches are
presented for several $q$ in the lower row of FIG. \ref{fig:a10alpha285},
which shows that the small branch, the tiny branch and the part of the large
branch that is far enough away from the critical line are radially stable. For
$q$ close enough to $q_{\text{cr}}^{\text{upp}}\left(  \alpha\right)  $, the
effective potentials of the large branch have negative regions, which occurs
away from the horizon. So the analysis of radial stability is inconclusive in
this case.

\begin{figure}[ptbh]
\begin{center}
\includegraphics[width=0.48\textwidth]{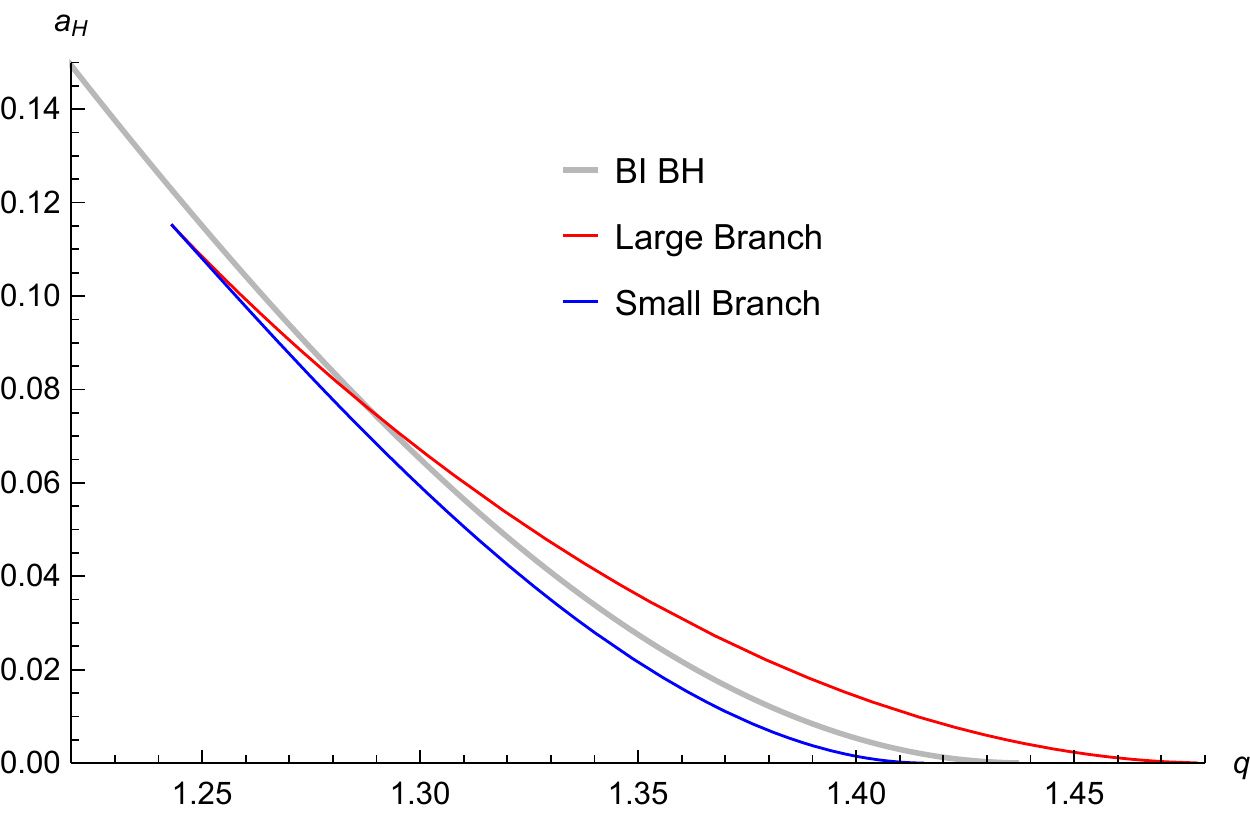}
\includegraphics[width=0.48\textwidth]{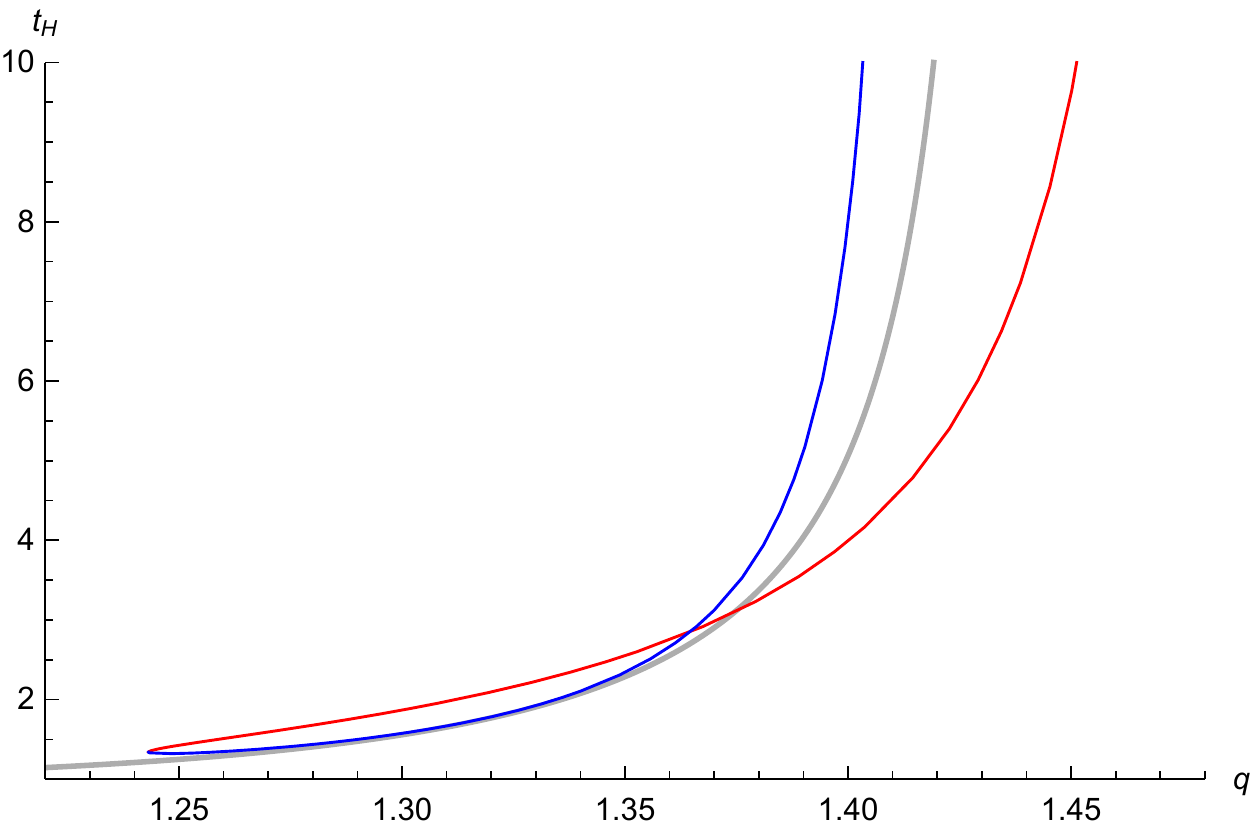}
\includegraphics[width=0.48\textwidth]{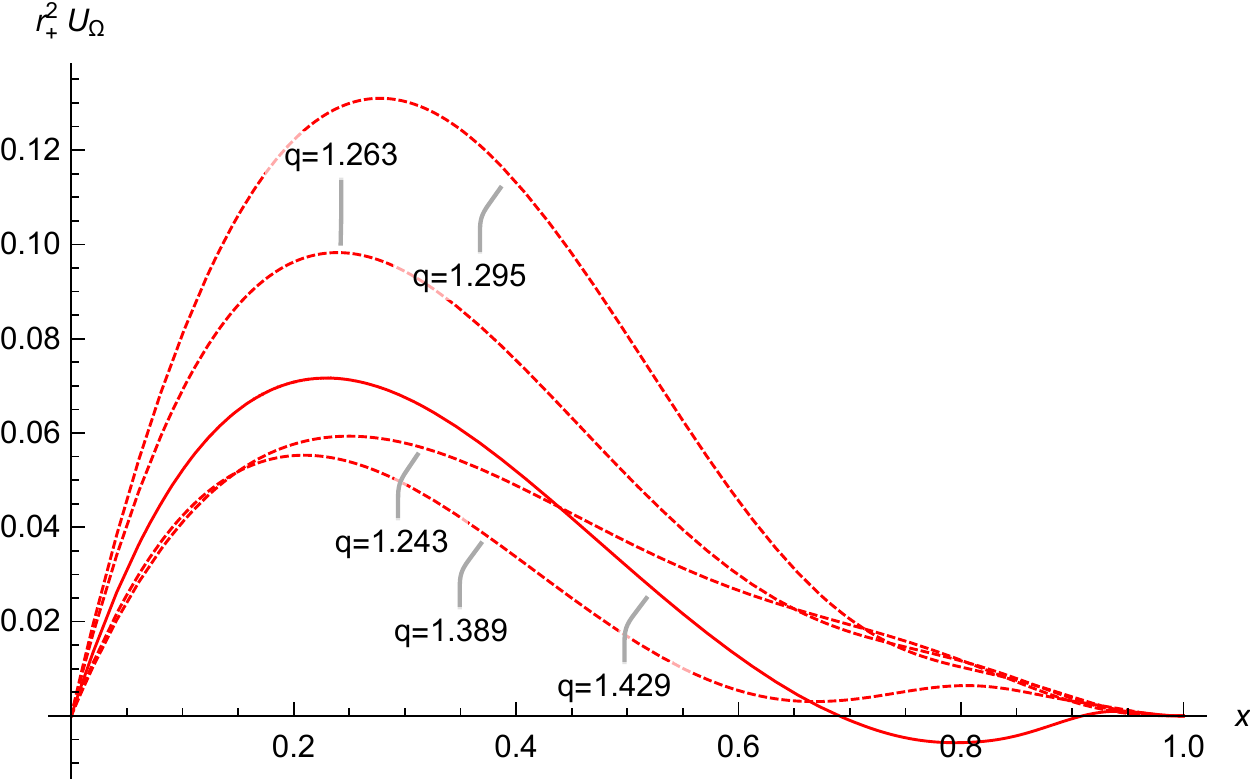}
\includegraphics[width=0.48\textwidth]{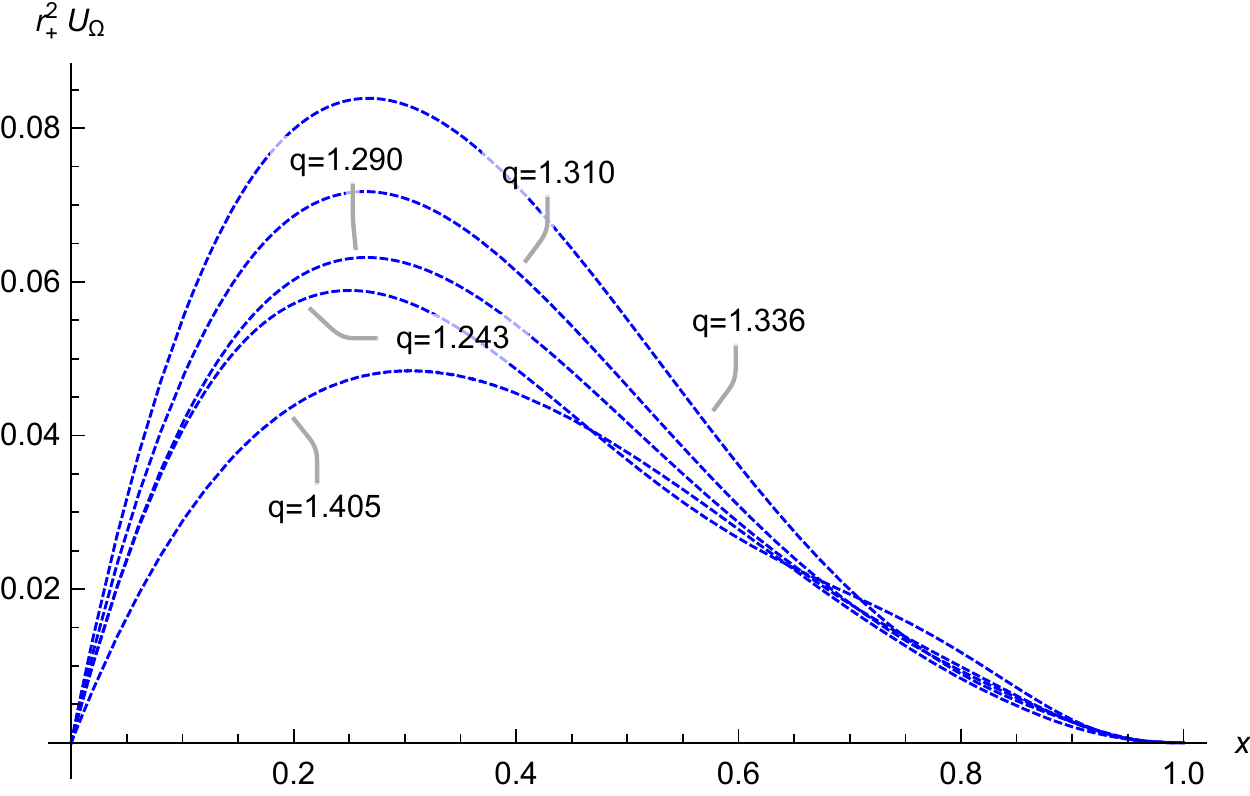}
\end{center}
\caption{{\footnotesize Thermodynamic preference, temperature and effective
potentials for the large (red lines) and small (blue lines) branches of
Schwarzschild-like BI black hole solutions with $\tilde{a}=10$ and
$\alpha=2.5$. \textbf{Upper Left: }Reduced area $a_{H}$ versus reduced charge
$q$ for BI black holes (a gray line) and the two branches of scalarized BI
black holes. For a given $q$, the small branch and the part of the large
branch with small $q$ have smaller area than BI black holes. When $q$ is large
enough, the large branch is entropically preferred. \textbf{Upper Right:
}Reduced temperature $t_{H}$ versus reduced charge $q$ for BI black holes and
the two branches of scalarized BI black holes. The large branch is hotter
(colder) than the small branch for small (large) values of $q$. \textbf{Lower}%
: Effective potentials for the two branches of scalarized BI black holes with
various values of $q$. The small branch and the large branch with small $q$
are stable against radial perturbations. However for large $q$, the stability
of the large branch is inconclusive.}}%
\label{fig:a10alpha25}%
\end{figure}

When $1.959\lesssim\alpha\lesssim2.846$, the bifurcation line ceases to exist,
and two branches of scalarized black holes, namely large branch and small
branch, would emerge and bifurcate from the existence line with the minimum
$q$ of the domain of existence. The large branch is the reminiscent of the
large branch in the $2.846\lesssim\alpha\lesssim3.138$ case, and also ends at
the upper critical line. On the other hand, the small branch is the
reminiscent of the small and tiny branches in the $2.846\lesssim\alpha
\lesssim3.138$ case, and terminates at the lower critical line or other
existence line. To better understand the two branch structure of the
scalarized solutions, we will consider scalarized solutions with $\alpha=2.5$
and $2.2$.

In FIG. \ref{fig:a10alpha25}, we plot $a_{H}$ versus $q$, $t_{H}$ versus $q$
and effective potentials for scalarized black hole solutions with $\alpha
=2.5$. The upper left panel justifies the terms for the large and small
branches, since it shows that the reduced area of the large branch is larger
than that of the small branch. The upper left panel also displays that the end
configurations of the two branches possess vanishing horizon area,
corresponding to the critical lines. In comparison with BI black holes, the
small branch also has smaller area. As $q$ increases from the existence line
toward the critical line, the reduced area of the large branch is initially
less and then becomes greater than that of BI black holes with the same $q$.
So the large branch is entropically preferred when $q$ is large enough.
Interestingly, the part of the large branch that is entropically disfavored
becomes larger in the $\alpha=2.5$ case than in the $\alpha=2.85$ case. The
upper right panel of FIG. \ref{fig:a10alpha25} exhibits that the temperatures
of the two branches are monotonically increasing functions of $q$.
Additionally, as $q$ grows, the large branch is first colder and then becomes
hotter than the small branch. In the lower row of FIG. \ref{fig:a10alpha25},
we present effective potentials for the two branches for several values of
$q$. Near the critical line, the effective potential of the large branch owns
a negative part, which means stability is inconclusive. However for the small
branch and the part of the large branch at a distance from the critical line,
the effective potentials are positive, which suggests that they are radially stable.

\begin{figure}[ptbh]
\begin{center}
\includegraphics[width=0.48\textwidth]{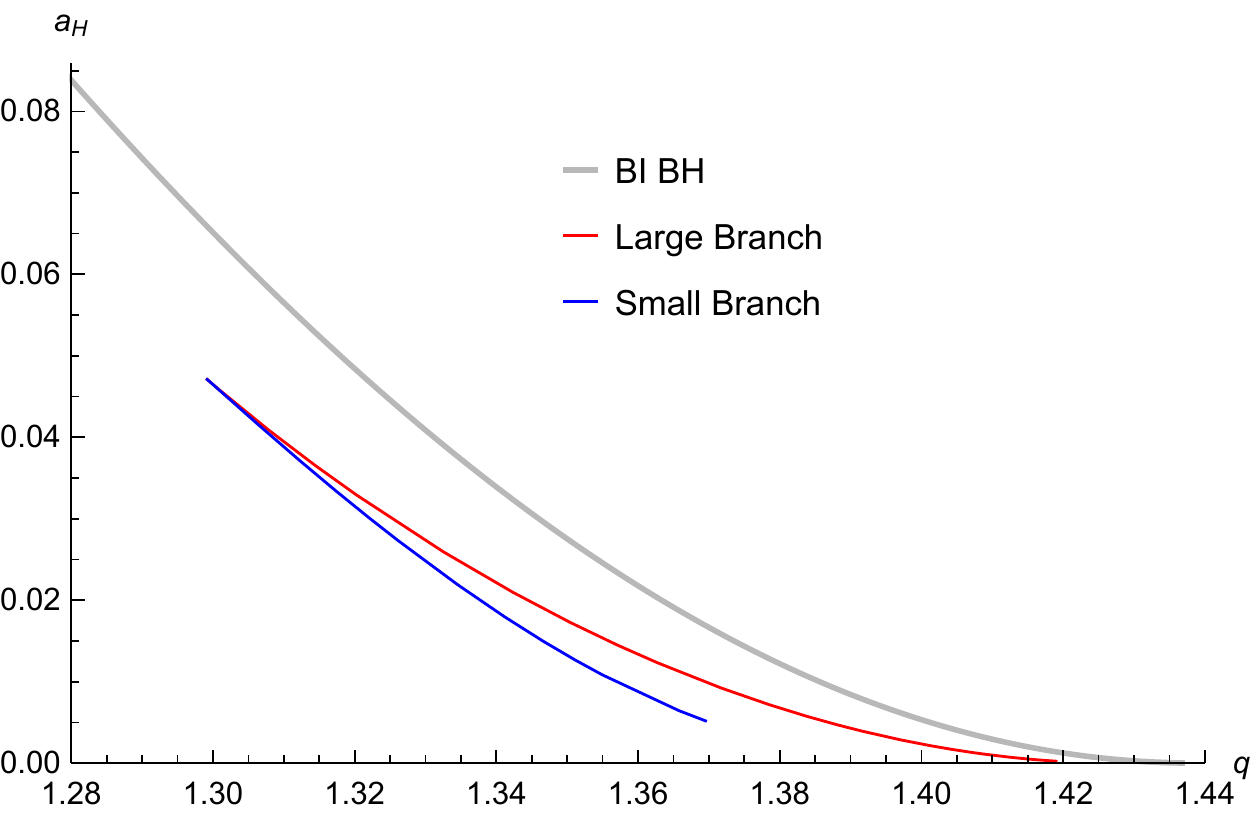}
\includegraphics[width=0.48\textwidth]{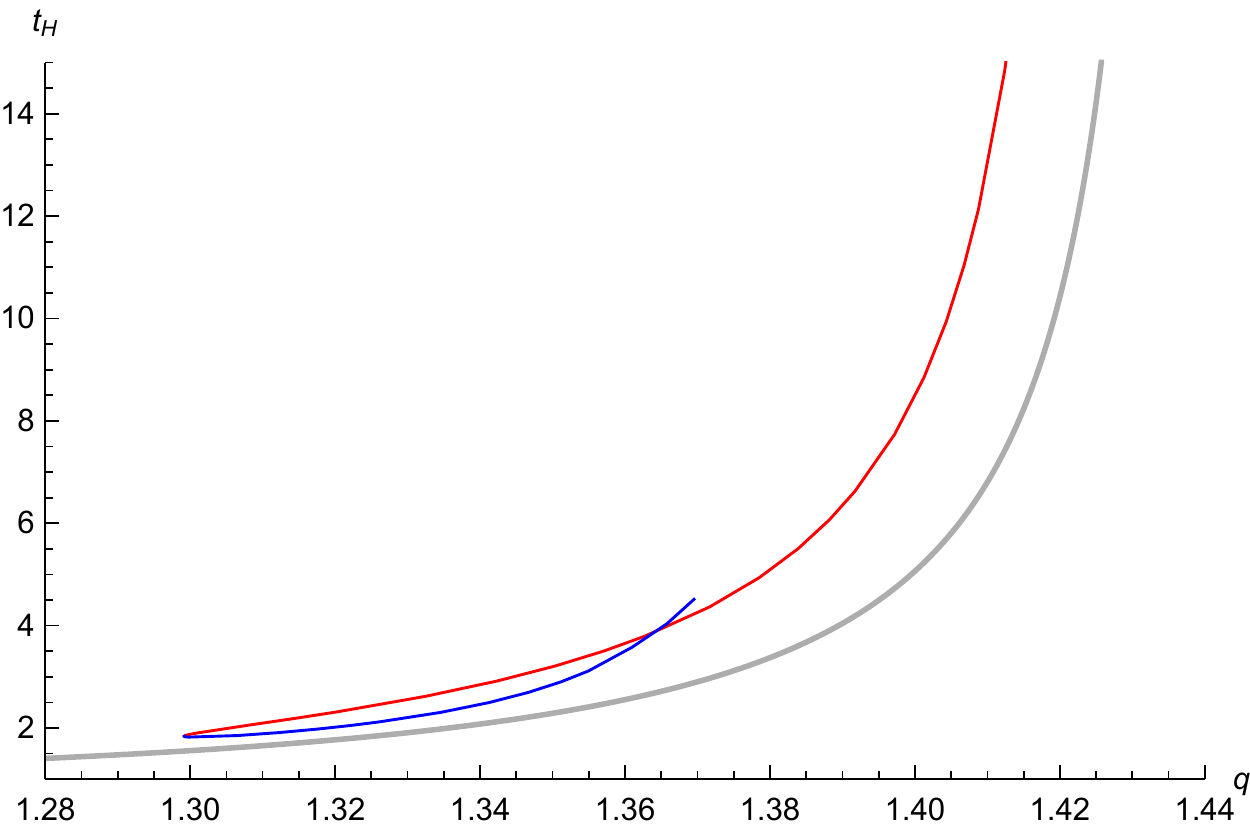}
\includegraphics[width=0.48\textwidth]{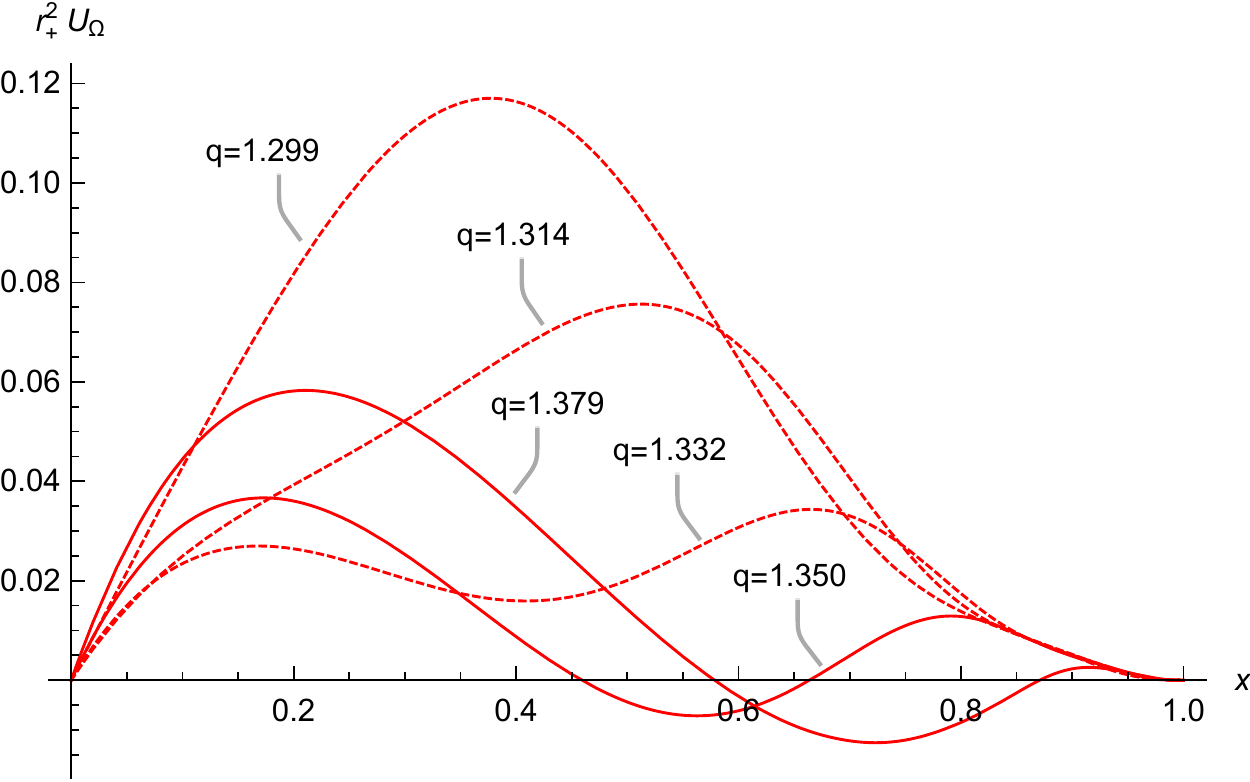}
\includegraphics[width=0.48\textwidth]{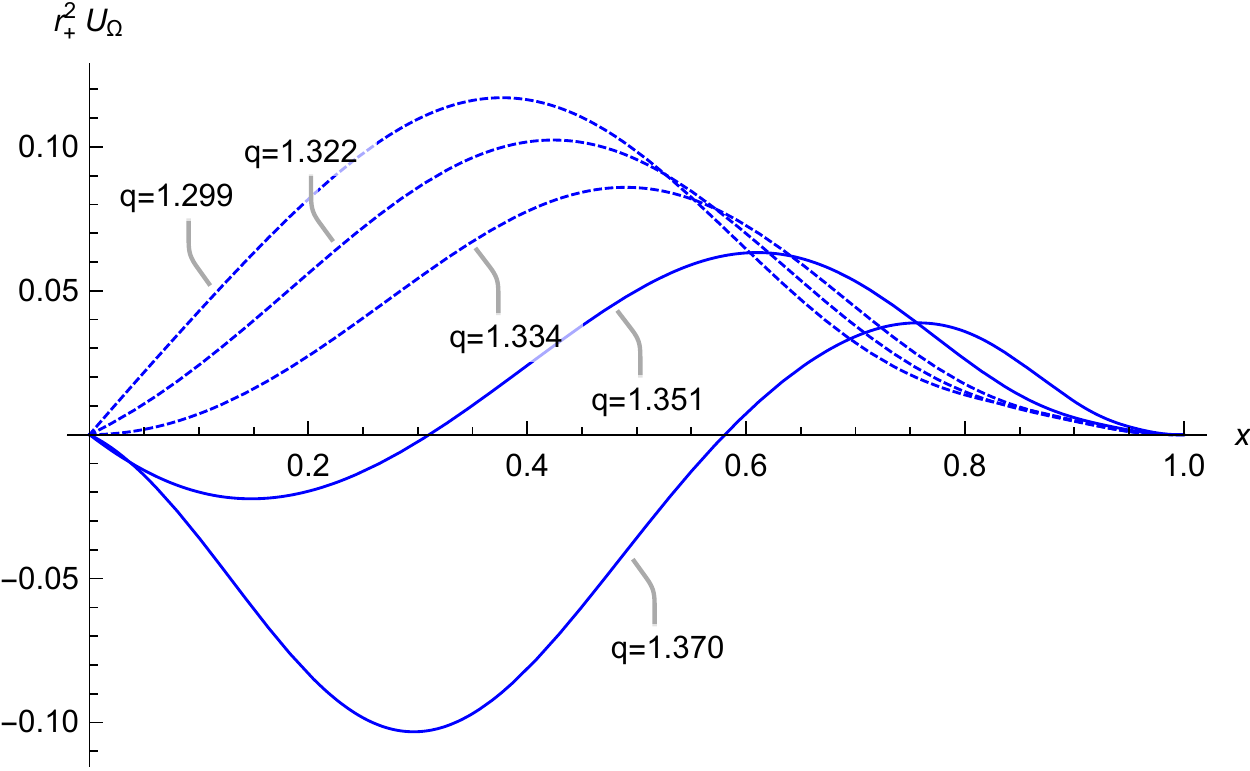}
\end{center}
\caption{{\footnotesize Thermodynamic preference, temperature and effective
potentials for the large (red lines) and small (blue lines) branches of
Schwarzschild-like BI black hole solutions with $\tilde{a}=10$ and
$\alpha=2.25$. \textbf{Upper Left: }Reduced area $a_{H}$ versus reduced charge
$q$ for BI black holes (a gray line) and the two branches of scalarized BI
black holes. For a given $q$, the small and large branches both have smaller
area than BI black holes, which means that the scalarized solutions are not
entropically preferred. \textbf{Upper Right: }Reduced temperature $t_{H}$
versus reduced charge $q$ for BI black holes and the two branches of
scalarized BI black holes. The scalarized solutions are hotter than BI black
holes, and the large branch is hotter (colder) than the small branch for small
(large) values of $q$. \textbf{Lower}: Effective potentials for the two
branches of scalarized BI black holes with various values of $q$. When $q$ is
large, the stability of the scalarized solutions is inconclusive. For small
enough $q$, the scalarized solutions are stable against radial perturbations.}%
}%
\label{fig:a10alpha22}%
\end{figure}

In FIG. \ref{fig:a10alpha22}, we present $a_{H}$ versus $q$, $t_{H}$ versus
$q$ and effective potentials for scalarized black hole solutions with
$\alpha=2.2$. According to the left upper panel, the large and small branches
end at the critical and some existence lines, respectively. Moreover, it shows
that the two branches are entropically disfavored since they have smaller area
than BI black holes with the same $q$. From the upper right panel, one
observes that, except small part of the small branch near the end
configuration, the temperature of the large branch is higher than that of the
small branch. Effective potentials for the two branches are plotted in the
lower row of FIG. \ref{fig:a10alpha22}, which indicates that the two branches
can be stable against radial perturbations when $q$ is small enough.

\section{Discussion and Conclusion}

\label{Sec:DC}

In this paper, we studied scalarized black hole solutions in the EBIS model
with the non-minimal coupling function $f\left(  \phi\right)  =e^{\alpha
\phi^{2}}$, which is subclass IIA according to the classification suggested in
\cite{Astefanesei:2019pfq}. In subclass IIA, scalar-free solutions suffer from
a tachyonic instability, and scalarized solutions can bifurcate from the
scalar-free solutions on bifurcation lines, which consists of zero modes of
the scalar-free solutions. In the EBIS model, scalar-free solutions are BI
black holes given in eqn. $\left(  \ref{eq:BIBH}\right)  $, which have been
shown to possess two types of solutions, i.e., RN-like and Schwarzschild-like
BI black holes. In the $\alpha$-$q$ plane, bifurcation lines of RN-like BI and
RN black holes are very similar (see FIG. \ref{fig:DoEa3}), whereas
bifurcation lines of Schwarzschild-like BI and RN black holes have quite
different behavior in small $\alpha$ regime (see FIG. \ref{fig:DoEa10}).
Specifically, the bifurcation line of Schwarzschild-like BI black holes is a
double-valued function of $\alpha$ in some region of the parameter space.

To gain insight into scalarized solutions, the domains of existence,
thermodynamic preference, radial stability and temperature of these solutions
were numerically investigated for the RN-like case with $\tilde{a}=3$ and the
Schwarzschild-like case with $\tilde{a}=10$, respectively. For the scalarized
RN-like black hole solutions, the domain of existence is bounded by the
existence and critical lines, which resembles the EMS models very closely (see
FIG. \ref{fig:DoEa3}). Moreover, we found that the scalarized RN-like black
hole solutions have only one branch, which is always entropically preferred
over the BI black holes with the same $q$, and almost radially stable except
for the parameter region close to the critical line in the small $\alpha$
regime (see FIG. \ref{fig:DoEa3}). These observations imply that the
scalarized solution are candidate endpoints of the evolution of unstable
RN-like BI black holes.

On the other hand, the domain of existence for the scalarized
Schwarzschild-like black hole solutions is different with the existence of a
new type of boundary, where scalarized solutions with a non-trivial scalar
field profile exist, and a new parameter region, where two branches of
solutions coexist (see FIG. \ref{fig:DoEa10}). When $\alpha$ is large enough,
the scalarized solutions have been shown to be entropically favoured over
comparable BI black holes and stable against radial perturbations (see FIG.
\ref{fig:a10alpha10}). In some small $\alpha$ regime, scalarized solutions
have three branches, which are named as the large, small and tiny branches,
respectively, according to the black hole reduced areas (see FIG.
\ref{fig:a10alpha285}). The small and tiny branches bifurcate from zero modes
of BI black holes, whereas the large branch does not connect with BI black
holes. Except small parameter region near the existence line, the large branch
of the solutions is always entropically favoured over the small and tiny
branches and comparable BI black holes. In addition, the large branch is
radially stable in the parameter region far away enough from the critical
line. Further decreasing $\alpha$, we found the small and tiny branches can
join together to form a new small branch of solutions, which becomes
disconnected from BI black holes (see FIG. \ref{fig:a10alpha25}). The large
branch is entropically favoured over comparable BI black holes when it is far
away enough from the existence line. The parameter region that the large
branch is entropically disfavored grows as $\alpha$ decreases. The large
branch is also radially stable when it is at a distance from the critical
line. Finally, there exists some smaller $\alpha$, for which BI black holes
are entropically favoured over the large and small branches of scalarized
solutions (see FIG. \ref{fig:a10alpha22}).

In this paper, we have presented preliminary results of the EBIS model with
the exponential coupling and observed that effects of NLED can play a
significant role in certain parameters. Let us conclude with some comments
regarding potential future research directions. It would be interesting to
study excited scalarized solutions since only the fundamental state was
considered in this paper. For the EMS model, it has been shown that the
fundamental state is entropically preferred over excited states
\cite{Herdeiro:2018wub}. Apart from the exponential coupling, studying other
coupling functions, e.g., a subclass IIB quartic coupling function, can reveal
new phenomena. Richer structure, e.g., hot and cold branches, have been
observed in the EMS model of subclass IIB \cite{Blazquez-Salcedo:2020nhs}.
Finally, obtaining rotating scalarized BI black holes and fully non-linear
evolutions of the EBIS model are always desirable.

\begin{acknowledgments}
We are grateful to Shuxuan Ying and Zhipeng Zhang for useful discussions and
valuable comments. This work is supported in part by NSFC (Grant No. 11875196,
11375121, 11947225 and 11005016).
\end{acknowledgments}

\bibliographystyle{unsrturl}
\bibliography{ref}

\end{document}